\documentclass[twocolumn,showpacs,preprintnumbers,amsmath,amssymb]{revtex4}%
\usepackage{amsfonts}
\usepackage{amsmath}
\usepackage{amssymb}
\usepackage{graphicx}
\usepackage{subfigure}%
\begin{document}
\preprint{ }
\title[ ]{{\large A study of local approximations for polarization potentials}}
\author{W.H.Z. C\'{a}rdenas$^{1}$}
\author{M.S. Hussein$^{1,2}$}
\thanks{Martin Gutzwiller Fellow, 2007/2008.}
\affiliation{$^{1}$Instituto de F\'{\i}sica, Universidade de S\~{a}o Paulo, C.P. 66318,
05389-970, S\~{a}o Paulo, Brazil}
\affiliation{$^{2}$Max-Planck-Institut f\"ur Physik komplexer Systeme}
\affiliation{N\"othnitzer Stra$\beta$e 38, D-01187 Dresden, Germany}
\author{L.F. Canto}
\affiliation{Instituto de F\'{\i}sica, Universidade Federal do Rio de Janeiro, C.P. 68528,
21941-972 Rio de Janeiro, Brazil}
\author{J. Lubian}
\affiliation{Instituto de F\'{\i}sica, Universidade Federal Fluminense, Av. Litor\^{a}nea
S-N, 24210-340 Niter\'{o}i, RJ, Brazil.}
\date{today}
\date{}
\date{}
\date{}
\date{}
\date{}
\date{}
\date{}
\date{}
\date{today}
\date{}
\date{}
\date{}
\date{}
\date{}
\date{}
\date{}
\date{}
\date{today}
\date{}
\date{}
\date{}
\date{}
\date{}
\date{}
\date{}
\date{}
\date{today}
\date{}
\date{}
\date{}
\date{}
\date{}
\date{}
\date{}
\date{}
\date{today}
\date{}
\date{}
\date{}
\date{}
\date{}
\date{}
\date{}
\date{}
\date{today}
\date{}
\date{}
\date{}
\date{}
\date{}
\date{}
\date{}
\date{}
\date{today}
\date{}
\date{}
\date{}
\date{}
\date{}
\date{}
\date{}
\date{}
\date{today}
\date{}
\date{}
\date{}
\date{}
\date{}
\date{}
\date{}
\date{}
\date{today}
\date{}
\date{}
\date{}
\date{}
\date{}
\date{}
\date{}
\date{}
\date{today}
\date{}
\date{}
\date{}
\date{}
\date{}
\date{}
\date{}
\date{}
\date{today}
\date{}
\date{}
\date{}
\date{}
\date{}
\date{}
\date{}
\date{}
\date{today}
\date{}
\date{}
\date{}
\date{}
\date{}
\date{}
\date{}
\date{}
\date{today}
\date{}
\date{}
\date{}
\date{}
\date{}
\date{}
\date{}
\date{}
\date{today}
\date{}
\date{}
\date{}
\date{}
\date{}
\date{}
\date{}
\date{}
\date{today}
\date{}
\date{}
\date{}
\date{}
\date{}
\date{}
\date{}
\date{}
\date{today}
\date{}
\date{}
\date{}
\date{}
\date{}
\date{}
\date{}
\date{}
\date{today}
\date{}
\date{}
\date{}
\date{}
\date{}
\date{}
\date{}
\date{}
\date{today}
\date{}
\date{}
\date{}
\date{}
\date{}
\date{}
\date{}
\date{}
\date{today}
\date{}
\date{}
\date{}
\date{}
\date{}
\date{}
\date{}
\date{}
\date{today}
\date{}
\date{}
\date{}
\date{}
\date{}
\date{}
\date{}
\date{}
\date{today}
\date{}
\date{}
\date{}
\date{}
\date{}
\date{}
\date{}
\date{}
\date{today}
\date{}
\date{}
\date{}
\date{}
\date{}
\date{}
\date{}
\date{}
\date{today}
\date{}
\date{}
\date{}
\date{}
\date{}
\date{}
\date{}
\date{}
\date{today}
\date{}
\date{}
\date{}
\date{}
\date{}
\date{}
\date{}
\date{}
\date{today}
\date{}
\date{}
\date{}
\date{}
\date{}
\date{}
\date{}
\date{}
\date{today}
\date{}
\date{}
\date{}
\date{}
\date{}
\date{}
\date{}
\date{}
\date{today}
\date{}
\date{}
\date{}
\date{}
\date{}
\date{}
\date{}
\date{}
\date{today}
\date{}
\date{}
\date{}
\date{}
\date{}
\date{}
\date{}
\date{}
\date{today}
\date{}
\date{}
\date{}
\date{}
\date{}
\date{}
\date{}
\date{}
\date{today}
\date{}
\date{}
\date{}
\date{}
\date{}
\date{}
\date{}
\date{}
\date{today}
\date{}
\date{}
\date{}
\date{}
\date{}
\date{}
\date{}
\date{}
\date{today}
\date{}
\date{}
\date{}
\date{}
\date{}
\date{}
\date{}
\date{}
\date{today}
\date{}
\date{}
\date{}
\date{}
\date{}
\date{}
\date{}
\date{}
\date{today}
\date{}
\date{}
\date{}
\date{}
\date{}
\date{}
\date{}
\date{}
\date{today}
\date{}
\date{}
\date{}
\date{}
\date{}
\date{}
\date{}
\date{}
\date{today}
\date{}
\date{}
\date{}
\date{}
\date{}
\date{}
\date{}
\date{}
\date{today}
\date{}
\date{}
\date{}
\date{}
\date{}
\date{}
\date{}
\date{}
\date{today}
\date{}
\date{}
\date{}
\date{}
\date{}
\date{}
\date{}
\date{}
\date{today}
\date{}
\date{}
\date{}
\date{}
\date{}
\date{}
\date{}
\date{}
\date{today}
\date{}
\date{}
\date{}
\date{}
\date{}
\date{}
\date{}
\date{}
\date{today}
\date{}
\date{}
\date{}
\date{}
\date{}
\date{}
\date{}
\date{}
\date{today}
\date{}
\date{}
\date{}
\date{}
\date{}
\date{}
\date{}
\date{}
\date{today}
\date{}
\date{}
\date{}
\date{}
\date{}
\date{}
\date{}
\date{}
\date{today}
\date{}
\date{}
\date{}
\date{}
\date{}
\date{}
\date{}
\date{}
\date{today}
\date{}
\date{}
\date{}
\date{}
\date{}
\date{}
\date{}
\date{}
\date{today}
\date{}
\date{}
\date{}
\date{}
\date{}
\date{}
\date{}
\date{}
\date{today}
\date{}
\date{}
\date{}
\date{}
\date{}
\date{}
\date{}
\date{}
\date{today}
\date{}
\date{}
\date{}
\date{}
\date{}
\date{}
\date{}
\date{}
\date{today}
\date{}
\date{}
\date{}
\date{}
\date{}
\date{}
\date{}
\date{}
\date{today}
\date{}
\date{}
\date{}
\date{}
\date{}
\date{}
\date{}
\date{}
\date{today}
\date{}
\date{}
\date{}
\date{}
\date{}
\date{}
\date{}
\date{}
\date{today}
\date{}
\date{}
\date{}
\date{}
\date{}
\date{}
\date{}
\date{}
\date{today}
\date{}
\date{}
\date{}
\date{}
\date{}
\date{}
\date{}
\date{}
\date{today}
\date{}
\date{}
\date{}
\date{}
\date{}
\date{}
\date{}
\date{}
\date{today}
\date{}
\date{}
\date{}
\date{}
\date{}
\date{}
\date{}
\date{}
\date{today}
\date{}
\date{}
\date{}
\date{}
\date{}
\date{}
\date{}
\date{}
\date{today}
\date{}
\date{}
\date{}
\date{}
\date{}
\date{}
\date{}
\date{}
\date{today}
\date{}
\date{}
\date{}
\date{}
\date{}
\date{}
\date{}
\date{}
\date{today}
\date{}
\date{}
\date{}
\date{}
\date{}
\date{}
\date{}
\date{}
\date{today}
\date{}
\date{}
\date{}
\date{}
\date{}
\date{}
\date{}
\date{}
\date{today}
\date{}
\date{}
\date{}
\date{}
\date{}
\date{}
\date{}
\date{}
\date{today}
\date{}
\date{}
\date{}
\date{}
\date{}
\date{}
\date{}
\date{}
\date{today}
\date{}
\date{}
\date{}
\date{}
\date{}
\date{}
\date{}
\date{}
\date{today}
\date{}
\date{}
\date{}
\date{}
\date{}
\date{}
\date{}
\date{}
\date{today}
\date{}
\date{}
\date{}
\date{}
\date{}
\date{}
\date{}
\date{}
\date{today}
\date{}
\date{}
\date{}
\date{}
\date{}
\date{}
\date{}
\date{}
\date{today}
\date{}
\date{}
\date{}
\date{}
\date{}
\date{}
\date{}
\date{}
\date{today}
\date{}
\date{}
\date{}
\date{}
\date{}
\date{}
\date{}
\date{}
\date{today}
\date{}
\date{}
\date{}
\date{}
\date{}
\date{}
\date{}
\date{}
\date{today}
\date{}
\date{}
\date{}
\date{}
\date{}
\date{}
\date{}
\date{}
\date{today}
\date{}
\date{}
\date{}
\date{}
\date{}
\date{}
\date{}
\date{}
\date{today}
\date{}
\date{}
\date{}
\date{}
\date{}
\date{}
\date{}
\date{}
\date{today}
\date{}
\date{}
\date{}
\date{}
\date{}
\date{}
\date{}
\date{}
\date{today}
\date{}
\date{}
\date{}
\date{}
\date{}
\date{}
\date{}
\date{}
\date{today}
\date{}
\date{}
\date{}
\date{}
\date{}
\date{}
\date{}
\date{}
\date{today}
\date{}
\date{}
\date{}
\date{}
\date{}
\date{}
\date{}
\date{}
\date{today}
\date{}
\date{}
\date{}
\date{}
\date{}
\date{}
\date{}
\date{}
\date{today}
\date{}
\date{}
\date{}
\date{}
\date{}
\date{}
\date{}
\date{}
\date{today}
\date{}
\date{}
\date{}
\date{}
\date{}
\date{}
\date{}
\date{}
\date{today}
\date{}
\date{}
\date{}
\date{}
\date{}
\date{}
\date{}
\date{}
\date{today}
\date{}
\date{}
\date{}
\date{}
\date{}
\date{}
\date{}
\date{}

\begin{abstract}
We discuss the derivation of an equivalent \textit{l}-independent polarization
potential for use in the optical Schr\"{o}dinger equation that describes the
elastic scattering of heavy ions. Three diffferent methods are used for this
purpose. Application of our theory to the low energy scattering of the halo
nucleus $^{11}$Li from a $^{12}$C target is made. It is found that the notion
of \ \textit{l}-independent polarization potential has some validity but can
not be a good substitute for the \textit{l}-dependent local equivalent
Feshbach polarization potential.

\end{abstract}
\volumeyear{ }
\volumenumber{ }
\issuenumber{ }
\eid{ }
\date{today}
\date{}
\date{}
\date{}
\date{}
\date{}
\date{}
\date{}
\date{}
\date{today}
\date{}
\date{}
\date{}
\date{}
\date{}
\date{}
\date{}
\date{}
\date{today}
\date{}
\date{}
\date{}
\date{}
\date{}
\date{}
\date{}
\date{}
\date{today}
\date{}
\date{}
\date{}
\date{}
\date{}
\date{}
\date{}
\date{}
\date{today}
\date{}
\date{}
\date{}
\date{}
\date{}
\date{}
\date{}
\date{}
\date{today}
\date{}
\date{}
\date{}
\date{}
\date{}
\date{}
\date{}
\date{}
\date{today}
\date{}
\date{}
\date{}
\date{}
\date{}
\date{}
\date{}
\date{}
\date{today}
\date{}
\date{}
\date{}
\date{}
\date{}
\date{}
\date{}
\date{}
\date{today}
\date{}
\date{}
\date{}
\date{}
\date{}
\date{}
\date{}
\date{}
\date{today}
\date{}
\date{}
\date{}
\date{}
\date{}
\date{}
\date{}
\date{}
\date{today}
\date{}
\date{}
\date{}
\date{}
\date{}
\date{}
\date{}
\date{}
\date{today}
\date{}
\date{}
\date{}
\date{}
\date{}
\date{}
\date{}
\date{}
\date{today}
\date{}
\date{}
\date{}
\date{}
\date{}
\date{}
\date{}
\date{}
\date{today}
\date{}
\date{}
\date{}
\date{}
\date{}
\date{}
\date{}
\date{}
\date{today}
\date{}
\date{}
\date{}
\date{}
\date{}
\date{}
\date{}
\date{}
\date{today}
\date{}
\date{}
\date{}
\date{}
\date{}
\date{}
\date{}
\date{}
\date{today}
\date{}
\date{}
\date{}
\date{}
\date{}
\date{}
\date{}
\date{}
\date{today}
\date{}
\date{}
\date{}
\date{}
\date{}
\date{}
\date{}
\date{}
\date{today}
\date{}
\date{}
\date{}
\date{}
\date{}
\date{}
\date{}
\date{}
\date{today}
\date{}
\date{}
\date{}
\date{}
\date{}
\date{}
\date{}
\date{}
\date{today}
\date{}
\date{}
\date{}
\date{}
\date{}
\date{}
\date{}
\date{}
\date{today}
\date{}
\date{}
\date{}
\date{}
\date{}
\date{}
\date{}
\date{}
\date{today}
\date{}
\date{}
\date{}
\date{}
\date{}
\date{}
\date{}
\date{}
\date{today}
\date{}
\date{}
\date{}
\date{}
\date{}
\date{}
\date{}
\date{}
\date{today}
\date{}
\date{}
\date{}
\date{}
\date{}
\date{}
\date{}
\date{}
\date{today}
\date{}
\date{}
\date{}
\date{}
\date{}
\date{}
\date{}
\date{}
\date{today}
\date{}
\date{}
\date{}
\date{}
\date{}
\date{}
\date{}
\date{}
\date{today}
\date{}
\date{}
\date{}
\date{}
\date{}
\date{}
\date{}
\date{}
\date{today}
\date{}
\date{}
\date{}
\date{}
\date{}
\date{}
\date{}
\date{}
\date{today}
\date{}
\date{}
\date{}
\date{}
\date{}
\date{}
\date{}
\date{}
\date{today}
\date{}
\date{}
\date{}
\date{}
\date{}
\date{}
\date{}
\date{}
\date{today}
\date{}
\date{}
\date{}
\date{}
\date{}
\date{}
\date{}
\date{}
\date{today}
\date{}
\date{}
\date{}
\date{}
\date{}
\date{}
\date{}
\date{}
\date{today}
\date{}
\date{}
\date{}
\date{}
\date{}
\date{}
\date{}
\date{}
\date{today}
\date{}
\date{}
\date{}
\date{}
\date{}
\date{}
\date{}
\date{}
\date{today}
\date{}
\date{}
\date{}
\date{}
\date{}
\date{}
\date{}
\date{}
\date{today}
\date{}
\date{}
\date{}
\date{}
\date{}
\date{}
\date{}
\date{}
\date{today}
\date{}
\date{}
\date{}
\date{}
\date{}
\date{}
\date{}
\date{}
\date{today}
\date{}
\date{}
\date{}
\date{}
\date{}
\date{}
\date{}
\date{}
\date{today}
\date{}
\date{}
\date{}
\date{}
\date{}
\date{}
\date{}
\date{}
\date{today}
\date{}
\date{}
\date{}
\date{}
\date{}
\date{}
\date{}
\date{}
\date{today}
\date{}
\date{}
\date{}
\date{}
\date{}
\date{}
\date{}
\date{}
\date{today}
\date{}
\date{}
\date{}
\date{}
\date{}
\date{}
\date{}
\date{}
\date{today}
\date{}
\date{}
\date{}
\date{}
\date{}
\date{}
\date{}
\date{}
\date{today}
\date{}
\date{}
\date{}
\date{}
\date{}
\date{}
\date{}
\date{}
\date{today}
\date{}
\date{}
\date{}
\date{}
\date{}
\date{}
\date{}
\date{}
\date{today}
\date{}
\date{}
\date{}
\date{}
\date{}
\date{}
\date{}
\date{}
\date{today}
\date{}
\date{}
\date{}
\date{}
\date{}
\date{}
\date{}
\date{}
\date{today}
\date{}
\date{}
\date{}
\date{}
\date{}
\date{}
\date{}
\date{}
\date{today}
\date{}
\date{}
\date{}
\date{}
\date{}
\date{}
\date{}
\date{}
\date{today}
\date{}
\date{}
\date{}
\date{}
\date{}
\date{}
\date{}
\date{}
\date{today}
\date{}
\date{}
\date{}
\date{}
\date{}
\date{}
\date{}
\date{}
\date{today}
\date{}
\date{}
\date{}
\date{}
\date{}
\date{}
\date{}
\date{}
\date{today}
\date{}
\date{}
\date{}
\date{}
\date{}
\date{}
\date{}
\date{}
\date{today}
\date{}
\date{}
\date{}
\date{}
\date{}
\date{}
\date{}
\date{}
\date{today}
\date{}
\date{}
\date{}
\date{}
\date{}
\date{}
\date{}
\date{}
\date{today}
\date{}
\date{}
\date{}
\date{}
\date{}
\date{}
\date{}
\date{}
\date{today}
\date{}
\date{}
\date{}
\date{}
\date{}
\date{}
\date{}
\date{}
\date{today}
\date{}
\date{}
\date{}
\date{}
\date{}
\date{}
\date{}
\date{}
\date{today}
\date{}
\date{}
\date{}
\date{}
\date{}
\date{}
\date{}
\date{}
\date{today}
\date{}
\date{}
\date{}
\date{}
\date{}
\date{}
\date{}
\date{}
\date{today}
\date{}
\date{}
\date{}
\date{}
\date{}
\date{}
\date{}
\date{}
\date{today}
\date{}
\date{}
\date{}
\date{}
\date{}
\date{}
\date{}
\date{}
\date{today}
\date{}
\date{}
\date{}
\date{}
\date{}
\date{}
\date{}
\date{}
\date{today}
\date{}
\date{}
\date{}
\date{}
\date{}
\date{}
\date{}
\date{}
\date{today}
\date{}
\date{}
\date{}
\date{}
\date{}
\date{}
\date{}
\date{}
\date{today}
\date{}
\date{}
\date{}
\date{}
\date{}
\date{}
\date{}
\date{}
\date{today}
\date{}
\date{}
\date{}
\date{}
\date{}
\date{}
\date{}
\date{}
\date{today}
\date{}
\date{}
\date{}
\date{}
\date{}
\date{}
\date{}
\date{}
\date{today}
\date{}
\date{}
\date{}
\date{}
\date{}
\date{}
\date{}
\date{}
\date{today}
\date{}
\date{}
\date{}
\date{}
\date{}
\date{}
\date{}
\date{}
\date{today}
\date{}
\date{}
\date{}
\date{}
\date{}
\date{}
\date{}
\date{}
\date{today}
\date{}
\date{}
\date{}
\date{}
\date{}
\date{}
\date{}
\date{}
\date{today}
\date{}
\date{}
\date{}
\date{}
\date{}
\date{}
\date{}
\date{}
\date{today}
\date{}
\date{}
\date{}
\date{}
\date{}
\date{}
\date{}
\date{}
\date{today}
\date{}
\date{}
\date{}
\date{}
\date{}
\date{}
\date{}
\date{}
\date{today}
\date{}
\date{}
\date{}
\date{}
\date{}
\date{}
\date{}
\date{}
\date{today}
\date{}
\date{}
\date{}
\date{}
\date{}
\date{}
\date{}
\date{}
\date{today}
\date{}
\date{}
\date{}
\date{}
\date{}
\date{}
\date{}
\date{}
\date{today}
\date{}
\date{}
\date{}
\date{}
\date{}
\date{}
\date{}
\date{}
\date{today}
\date{}
\date{}
\date{}
\date{}
\date{}
\date{}
\date{}
\date{}
\date{today}
\date{}
\date{}
\date{}
\date{}
\date{}
\date{}
\date{}
\date{}
\date{today}
\date{}
\date{}
\date{}
\date{}
\date{}
\date{}
\date{}
\date{}
\date{today}
\date{}
\date{}
\date{}
\date{}
\date{}
\date{}
\date{}
\date{}
\date{today}
\date{}
\date{}
\date{}
\date{}
\date{}
\date{}
\date{}
\date{}
\date{today}
\date{}
\date{}
\date{}
\date{}
\date{}
\date{}
\date{}
\date{}
\date{today}
\date{}
\date{}
\date{}
\date{}
\date{}
\date{}
\date{}
\date{}
\date{today}
\date{}
\date{}
\date{}
\date{}
\date{}
\date{}
\date{}
\date{}
\date{today}
\date{}
\date{}
\date{}
\date{}
\date{}
\date{}
\date{}
\date{}
\date{today}
\date{}
\date{}
\date{}
\date{}
\date{}
\date{}
\date{}
\date{}
\date{today}
\date{}
\date{}
\date{}
\date{}
\date{}
\date{}
\date{}
\date{}
\date{today}
\date{}
\date{}
\date{}
\date{}
\date{}
\date{}
\date{}
\date{}
\date{today}
\date{}
\date{}
\date{}
\date{}
\date{}
\date{}
\date{}
\date{}
\date{today}
\date{}
\date{}
\date{}
\date{}
\date{}
\date{}
\date{}
\date{}
\date{today}
\date{}
\date{}
\date{}
\date{}
\date{}
\date{}
\date{}
\date{}
\date{today}
\date{}
\date{}
\date{}
\date{}
\date{}
\date{}
\date{}
\date{}
\date{today}
\date{}
\date{}
\date{}
\date{}
\date{}
\date{}
\date{}
\date{}
\date{today}
\date{}
\date{}
\date{}
\date{}
\date{}
\date{}
\date{}
\date{}
\date{today}
\date{}
\date{}
\date{}
\date{}
\date{}
\date{}
\date{}
\date{}
\date{today}
\date{}
\date{}
\date{}
\date{}
\date{}
\date{}
\date{}
\date{}
\date{today}
\date{}
\date{}
\date{}
\date{}
\date{}
\date{}
\date{}
\date{}
\date{today}
\date{}
\date{}
\date{}
\date{}
\date{}
\date{}
\date{}
\date{}
\date{today}
\date{}
\date{}
\date{}
\date{}
\date{}
\date{}
\date{}
\date{}
\date{today}
\date{}
\date{}
\date{}
\date{}
\date{}
\date{}
\date{}
\date{}
\date{today}
\date{}
\date{}
\date{}
\date{}
\date{}
\date{}
\date{}
\date{}
\date{today}
\date{}
\date{}
\date{}
\date{}
\date{}
\date{}
\date{}
\date{}
\date{today}
\date{}
\date{}
\date{}
\date{}
\date{}
\date{}
\date{}
\date{}
\date{today}
\date{}
\date{}
\date{}
\date{}
\date{}
\date{}
\date{}
\date{}
\date{today}
\date{}
\date{}
\date{}
\date{}
\date{}
\date{}
\date{}
\date{}
\date{today}
\date{}
\date{}
\date{}
\date{}
\date{}
\date{}
\date{}
\date{}
\date{today}
\date{}
\date{}
\date{}
\date{}
\date{}
\date{}
\date{}
\date{}
\date{today}
\date{}
\date{}
\date{}
\date{}
\date{}
\date{}
\date{}
\date{}
\date{today}
\date{}
\date{}
\date{}
\date{}
\date{}
\date{}
\date{}
\date{}
\date{today}
\date{}
\date{}
\date{}
\date{}
\date{}
\date{}
\date{}
\date{}
\date{today}
\date{}
\date{}
\date{}
\date{}
\date{}
\date{}
\date{}
\date{}
\date{today}
\date{}
\date{}
\date{}
\date{}
\date{}
\date{}
\date{}
\date{}
\date{today}
\date{}
\date{}
\date{}
\date{}
\date{}
\date{}
\date{}
\date{}
\date{today}
\date{}
\date{}
\date{}
\date{}
\date{}
\date{}
\date{}
\date{}
\date{today}
\date{}
\date{}
\date{}
\date{}
\date{}
\date{}
\date{}
\date{}
\date{today}
\date{}
\date{}
\date{}
\date{}
\date{}
\date{}
\date{}
\date{}
\date{today}
\date{}
\date{}
\date{}
\date{}
\date{}
\date{}
\date{}
\date{}
\date{today}
\date{}
\date{}
\date{}
\date{}
\date{}
\date{}
\date{}
\date{}
\date{today}
\date{}
\date{}
\date{}
\date{}
\date{}
\date{}
\date{}
\date{}
\date{today}
\date{}
\date{}
\date{}
\date{}
\date{}
\date{}
\date{}
\date{}
\date{today}
\date{}
\date{}
\date{}
\date{}
\date{}
\date{}
\date{}
\date{}
\date{today}
\date{}
\date{}
\date{}
\date{}
\date{}
\date{}
\date{}
\date{}
\date{today}
\date{}
\date{}
\date{}
\date{}
\date{}
\date{}
\date{}
\date{}
\date{today}
\date{}
\date{}
\date{}
\date{}
\date{}
\date{}
\date{}
\date{}
\date{today}
\date{}
\date{}
\date{}
\date{}
\date{}
\date{}
\date{}
\date{}
\date{today}
\date{}
\date{}
\date{}
\date{}
\date{}
\date{}
\date{}
\date{}
\date{today}
\date{}
\date{}
\date{}
\date{}
\date{}
\date{}
\date{}
\date{}
\date{today}
\date{}
\date{}
\date{}
\date{}
\date{}
\date{}
\date{}
\date{}
\date{today}
\date{}
\date{}
\date{}
\date{}
\date{}
\date{}
\date{}
\date{}
\date{today}
\date{}
\date{}
\date{}
\date{}
\date{}
\date{}
\date{}
\date{}
\date{today}
\date{}
\date{}
\date{}
\date{}
\date{}
\date{}
\date{}
\date{}
\date{today}
\date{}
\date{}
\date{}
\date{}
\date{}
\date{}
\date{}
\date{}
\date{today}
\date{}
\date{}
\date{}
\date{}
\date{}
\date{}
\date{}
\date{}
\date{today}
\date{}
\date{}
\date{}
\date{}
\date{}
\date{}
\date{}
\date{}
\date{today}
\date{}
\date{}
\date{}
\date{}
\date{}
\date{}
\date{}
\date{}
\date{today}
\date{}
\date{}
\date{}
\date{}
\date{}
\date{}
\date{}
\date{}
\date{today}
\date{}
\date{}
\date{}
\date{}
\date{}
\date{}
\date{}
\date{}
\date{today}
\date{}
\date{}
\date{}
\date{}
\date{}
\date{}
\date{}
\date{}
\date{today}
\date{}
\date{}
\date{}
\date{}
\date{}
\date{}
\date{}
\date{}
\date{today}
\date{}
\date{}
\date{}
\date{}
\date{}
\date{}
\date{}
\date{}
\date{today}
\date{}
\date{}
\date{}
\date{}
\date{}
\date{}
\date{}
\date{}
\date{today}
\date{}
\date{}
\date{}
\date{}
\date{}
\date{}
\date{}
\date{}
\date{today}
\date{}
\date{}
\date{}
\date{}
\date{}
\date{}
\date{}
\date{}
\date{today}
\date{}
\date{}
\date{}
\date{}
\date{}
\date{}
\date{}
\date{}
\date{today}
\date{}
\date{}
\date{}
\date{}
\date{}
\date{}
\date{}
\date{}
\date{today}
\date{}
\date{}
\date{}
\date{}
\date{}
\date{}
\date{}
\date{}
\date{today}
\date{}
\date{}
\date{}
\date{}
\date{}
\date{}
\date{}
\date{}
\date{today}
\date{}
\date{}
\date{}
\date{}
\date{}
\date{}
\date{}
\date{}
\date{today}
\date{}
\date{}
\date{}
\date{}
\date{}
\date{}
\date{}
\date{}
\date{today}
\date{}
\date{}
\date{}
\date{}
\date{}
\date{}
\date{}
\date{}
\date{today}
\date{}
\date{}
\date{}
\date{}
\date{}
\date{}
\date{}
\date{}
\date{today}
\date{}
\date{}
\date{}
\date{}
\date{}
\date{}
\date{}
\date{}
\date{today}
\date{}
\date{}
\date{}
\date{}
\date{}
\date{}
\date{}
\date{}
\date{today}
\date{}
\date{}
\date{}
\date{}
\date{}
\date{}
\date{}
\date{}
\date{today}
\date{}
\date{}
\date{}
\date{}
\date{}
\date{}
\date{}
\date{}
\date{today}
\date{}
\date{}
\date{}
\date{}
\date{}
\date{}
\date{}
\date{}
\date{today}
\date{}
\date{}
\date{}
\date{}
\date{}
\date{}
\date{}
\date{}
\date{today}
\date{}
\date{}
\date{}
\date{}
\date{}
\date{}
\date{}
\date{}
\date{today}
\date{}
\date{}
\date{}
\date{}
\date{}
\date{}
\date{}
\date{}
\date{today}
\date{}
\date{}
\date{}
\date{}
\date{}
\date{}
\date{}
\date{}
\date{today}
\date{}
\date{}
\date{}
\date{}
\date{}
\date{}
\date{}
\date{}
\date{today}
\date{}
\date{}
\date{}
\date{}
\date{}
\date{}
\date{}
\date{}
\date{today}
\date{}
\date{}
\date{}
\date{}
\date{}
\date{}
\date{}
\date{}
\date{today}
\date{}
\date{}
\date{}
\date{}
\date{}
\date{}
\date{}
\date{}
\date{today}
\date{}
\date{}
\date{}
\date{}
\date{}
\date{}
\date{}
\date{}
\date{today}
\date{}
\date{}
\date{}
\date{}
\date{}
\date{}
\date{}
\date{}
\date{today}
\date{}
\date{}
\date{}
\date{}
\date{}
\date{}
\date{}
\date{}
\date{today}
\date{}
\date{}
\date{}
\date{}
\date{}
\date{}
\date{}
\date{}
\date{today}
\date{}
\date{}
\date{}
\date{}
\date{}
\date{}
\date{}
\date{}
\date{today}
\date{}
\date{}
\date{}
\date{}
\date{}
\date{}
\date{}
\date{}
\date{today}
\date{}
\date{}
\date{}
\date{}
\date{}
\date{}
\date{}
\date{}
\date{today}
\date{}
\date{}
\date{}
\date{}
\date{}
\date{}
\date{}
\date{}
\date{today}
\date{}
\date{}
\date{}
\date{}
\date{}
\date{}
\date{}
\date{}
\date{today}
\date{}
\date{}
\date{}
\date{}
\date{}
\date{}
\date{}
\date{}
\date{today}
\date{}
\date{}
\date{}
\date{}
\date{}
\date{}
\date{}
\date{}
\date{today}
\date{}
\date{}
\date{}
\date{}
\date{}
\date{}
\date{}
\date{}
\date{today}
\date{}
\date{}
\date{}
\date{}
\date{}
\date{}
\date{}
\date{}
\date{today}
\date{}
\date{}
\date{}
\date{}
\date{}
\date{}
\date{}
\date{}
\date{today}
\date{}
\date{}
\date{}
\date{}
\date{}
\date{}
\date{}
\date{}
\date{today}
\date{}
\date{}
\date{}
\date{}
\date{}
\date{}
\date{}
\date{}
\date{today}
\date{}
\date{}
\date{}
\date{}
\date{}
\date{}
\date{}
\date{}
\date{today}
\date{}
\date{}
\date{}
\date{}
\date{}
\date{}
\date{}
\date{}
\date{today}
\date{}
\date{}
\date{}
\date{}
\date{}
\date{}
\date{}
\date{}
\date{today}
\date{}
\date{}
\date{}
\date{}
\date{}
\date{}
\date{}
\date{}
\date{today}
\date{}
\date{}
\date{}
\date{}
\date{}
\date{}
\date{}
\date{}
\date{today}
\date{}
\date{}
\date{}
\date{}
\date{}
\date{}
\date{}
\date{}
\date{today}
\date{}
\date{}
\date{}
\date{}
\date{}
\date{}
\date{}
\date{}
\date{today}
\date{}
\date{}
\date{}
\date{}
\date{}
\date{}
\date{}
\date{}
\date{today}
\date{}
\date{}
\date{}
\date{}
\date{}
\date{}
\date{}
\date{}
\date{today}
\date{}
\date{}
\date{}
\date{}
\date{}
\date{}
\date{}
\date{}
\date{today}
\date{}
\date{}
\date{}
\date{}
\date{}
\date{}
\date{}
\date{}
\date{today}
\date{}
\date{}
\date{}
\date{}
\date{}
\date{}
\date{}
\date{}
\date{today}
\date{}
\date{}
\date{}
\date{}
\date{}
\date{}
\date{}
\date{}
\date{today}
\date{}
\date{}
\date{}
\date{}
\date{}
\date{}
\date{}
\date{}
\date{today}
\date{}
\date{}
\date{}
\date{}
\date{}
\date{}
\date{}
\date{}
\date{today}
\date{}
\date{}
\date{}
\date{}
\date{}
\date{}
\date{}
\date{}
\date{today}
\date{}
\date{}
\date{}
\date{}
\date{}
\date{}
\date{}
\date{}
\date{today}
\date{}
\date{}
\date{}
\date{}
\date{}
\date{}
\date{}
\date{}
\date{today}
\date{}
\date{}
\date{}
\date{}
\date{}
\date{}
\date{}
\date{}
\date{today}
\date{}
\date{}
\date{}
\date{}
\date{}
\date{}
\date{}
\date{}
\date{today}
\date{}
\date{}
\date{}
\date{}
\date{}
\date{}
\date{}
\date{}
\date{today}
\date{}
\date{}
\date{}
\date{}
\date{}
\date{}
\date{}
\date{}
\date{today}
\date{}
\date{}
\date{}
\date{}
\date{}
\date{}
\date{}
\date{}
\date{today}
\date{}
\date{}
\date{}
\date{}
\date{}
\date{}
\date{}
\date{}
\date{today}
\date{}
\date{}
\date{}
\date{}
\date{}
\date{}
\date{}
\date{}
\date{today}
\date{}
\date{}
\date{}
\date{}
\date{}
\date{}
\date{}
\date{}
\date{today}
\date{}
\date{}
\date{}
\date{}
\date{}
\date{}
\date{}
\date{}
\date{today}
\date{}
\date{}
\date{}
\date{}
\date{}
\date{}
\date{}
\date{}
\date{today}
\date{}
\date{}
\date{}
\date{}
\date{}
\date{}
\date{}
\date{}
\date{today}
\date{}
\date{}
\date{}
\date{}
\date{}
\date{}
\date{}
\date{}
\date{today}
\date{}
\date{}
\date{}
\date{}
\date{}
\date{}
\date{}
\date{}
\date{today}
\date{}
\date{}
\date{}
\date{}
\date{}
\date{}
\date{}
\date{}
\date{today}
\date{}
\date{}
\date{}
\date{}
\date{}
\date{}
\date{}
\date{}
\date{today}
\date{}
\date{}
\date{}
\date{}
\date{}
\date{}
\date{}
\date{}
\date{today}
\date{}
\date{}
\date{}
\date{}
\date{}
\date{}
\date{}
\date{}
\date{today}
\date{}
\date{}
\date{}
\date{}
\date{}
\date{}
\date{}
\date{}
\date{today}
\date{}
\date{}
\date{}
\date{}
\date{}
\date{}
\date{}
\date{}
\date{today}
\date{}
\date{}
\date{}
\date{}
\date{}
\date{}
\date{}
\date{}
\date{today}
\date{}
\date{}
\date{}
\date{}
\date{}
\date{}
\date{}
\date{}
\date{today}
\date{}
\date{}
\date{}
\date{}
\date{}
\date{}
\date{}
\date{}
\date{today}
\date{}
\date{}
\date{}
\date{}
\date{}
\date{}
\date{}
\date{}
\date{today}
\date{}
\date{}
\date{}
\date{}
\date{}
\date{}
\date{}
\date{}
\date{today}
\date{}
\date{}
\date{}
\date{}
\date{}
\date{}
\date{}
\date{}
\date{today}
\date{}
\date{}
\date{}
\date{}
\date{}
\date{}
\date{}
\date{}
\date{today}
\date{}
\date{}
\date{}
\date{}
\date{}
\date{}
\date{}
\date{}
\date{today}
\date{}
\date{}
\date{}
\date{}
\date{}
\date{}
\date{}
\date{}
\date{today}
\date{}
\date{}
\date{}
\date{}
\date{}
\date{}
\date{}
\date{}
\date{today}
\date{}
\date{}
\date{}
\date{}
\date{}
\date{}
\date{}
\date{}
\date{today}
\date{}
\date{}
\date{}
\date{}
\date{}
\date{}
\date{}
\date{}
\date{today}
\date{}
\date{}
\date{}
\date{}
\date{}
\date{}
\date{}
\date{}
\date{today}
\date{}
\date{}
\date{}
\date{}
\date{}
\date{}
\date{}
\date{}
\date{today}
\date{}
\date{}
\date{}
\date{}
\date{}
\date{}
\date{}
\date{}
\date{today}
\date{}
\date{}
\date{}
\date{}
\date{}
\date{}
\date{}
\date{}
\date{today}
\date{}
\date{}
\date{}
\date{}
\date{}
\date{}
\date{}
\date{}
\date{today}
\date{}
\date{}
\date{}
\date{}
\date{}
\date{}
\date{}
\date{}
\date{today}
\date{}
\date{}
\date{}
\date{}
\date{}
\date{}
\date{}
\date{}
\date{today}
\date{}
\date{}
\date{}
\date{}
\date{}
\date{}
\date{}
\date{}
\date{today}
\date{}
\date{}
\date{}
\date{}
\date{}
\date{}
\date{}
\date{}
\date{today}
\date{}
\date{}
\date{}
\date{}
\date{}
\date{}
\date{}
\date{}
\date{today}
\date{}
\date{}
\date{}
\date{}
\date{}
\date{}
\date{}
\date{}
\date{today}
\date{}
\date{}
\date{}
\date{}
\date{}
\date{}
\date{}
\date{}
\date{today}
\date{}
\date{}
\date{}
\date{}
\date{}
\date{}
\date{}
\date{}
\date{today}
\date{}
\date{}
\date{}
\date{}
\date{}
\date{}
\date{}
\date{}
\date{today}
\date{}
\date{}
\date{}
\date{}
\date{}
\date{}
\date{}
\date{}
\date{today}
\date{}
\date{}
\date{}
\date{}
\date{}
\date{}
\date{}
\date{}
\date{today}
\date{}
\date{}
\date{}
\date{}
\date{}
\date{}
\date{}
\date{}
\date{today}
\date{}
\date{}
\date{}
\date{}
\date{}
\date{}
\date{}
\date{}
\date{today}
\date{}
\date{}
\date{}
\date{}
\date{}
\date{}
\date{}
\date{}
\date{today}
\date{}
\date{}
\date{}
\date{}
\date{}
\date{}
\date{}
\date{}
\date{today}
\date{}
\date{}
\date{}
\date{}
\date{}
\date{}
\date{}
\date{}
\date{today}
\date{}
\date{}
\date{}
\date{}
\date{}
\date{}
\date{}
\date{}
\date{today}
\date{}
\date{}
\date{}
\date{}
\date{}
\date{}
\date{}
\date{}
\date{today}
\date{}
\date{}
\date{}
\date{}
\date{}
\date{}
\date{}
\date{}
\date{today}
\date{}
\date{}
\date{}
\date{}
\date{}
\date{}
\date{}
\date{}
\date{today}
\date{}
\date{}
\date{}
\date{}
\date{}
\date{}
\date{}
\date{}
\date{today}
\date{}
\date{}
\date{}
\date{}
\date{}
\date{}
\date{}
\date{}
\date{today}
\date{}
\date{}
\date{}
\date{}
\date{}
\date{}
\date{}
\date{}
\date{today}
\date{}
\date{}
\date{}
\date{}
\date{}
\date{}
\date{}
\date{}
\date{today}
\date{}
\date{}
\date{}
\date{}
\date{}
\date{}
\date{}
\date{}
\date{today}
\date{}
\date{}
\date{}
\date{}
\date{}
\date{}
\date{}
\date{}
\date{today}
\date{}
\date{}
\date{}
\date{}
\date{}
\date{}
\date{}
\date{}
\date{today}
\date{}
\date{}
\date{}
\date{}
\date{}
\date{}
\date{}
\date{}
\date{today}
\date{}
\date{}
\date{}
\date{}
\date{}
\date{}
\date{}
\date{}
\date{today}
\date{}
\date{}
\date{}
\date{}
\date{}
\date{}
\date{}
\date{}
\date{today}
\date{}
\date{}
\date{}
\date{}
\date{}
\date{}
\date{}
\date{}
\date{today}
\date{}
\date{}
\date{}
\date{}
\date{}
\date{}
\date{}
\date{}
\date{today}
\date{}
\date{}
\date{}
\date{}
\date{}
\date{}
\date{}
\date{}
\date{today}
\date{}
\date{}
\date{}
\date{}
\date{}
\date{}
\date{}
\date{}
\date{today}
\date{}
\date{}
\date{}
\date{}
\date{}
\date{}
\date{}
\date{}
\date{today}
\date{}
\date{}
\date{}
\date{}
\date{}
\date{}
\date{}
\date{}
\date{today}
\date{}
\date{}
\date{}
\date{}
\date{}
\date{}
\date{}
\date{}
\date{today}
\date{}
\date{}
\date{}
\date{}
\date{}
\date{}
\date{}
\date{}
\date{today}
\date{}
\date{}
\date{}
\date{}
\date{}
\date{}
\date{}
\date{}
\date{today}
\date{}
\date{}
\date{}
\date{}
\date{}
\date{}
\date{}
\date{}
\date{today}
\date{}
\date{}
\date{}
\date{}
\date{}
\date{}
\date{}
\date{}
\date{today}
\date{}
\date{}
\date{}
\date{}
\date{}
\date{}
\date{}
\date{}
\date{today}
\date{}
\date{}
\date{}
\date{}
\date{}
\date{}
\date{}
\date{}
\date{today}
\date{}
\date{}
\date{}
\date{}
\date{}
\date{}
\date{}
\date{}
\date{today}
\date{}
\date{}
\date{}
\date{}
\date{}
\date{}
\date{}
\date{}
\date{today}
\date{}
\date{}
\date{}
\date{}
\date{}
\date{}
\date{}
\date{}
\date{today}
\date{}
\date{}
\date{}
\date{}
\date{}
\date{}
\date{}
\date{}
\date{today}
\date{}
\date{}
\date{}
\date{}
\date{}
\date{}
\date{}
\date{}
\date{today}
\date{}
\date{}
\date{}
\date{}
\date{}
\date{}
\date{}
\date{}
\date{today}
\date{}
\date{}
\date{}
\date{}
\date{}
\date{}
\date{}
\date{}
\date{today}
\date{}
\date{}
\date{}
\date{}
\date{}
\date{}
\date{}
\date{}
\date{today}
\date{}
\date{}
\date{}
\date{}
\date{}
\date{}
\date{}
\date{}
\date{today}
\date{}
\date{}
\date{}
\date{}
\date{}
\date{}
\date{}
\date{}
\date{today}
\date{}
\date{}
\date{}
\date{}
\date{}
\date{}
\date{}
\date{}
\date{today}
\date{}
\date{}
\date{}
\date{}
\date{}
\date{}
\date{}
\date{}
\date{today}
\date{}
\date{}
\date{}
\date{}
\date{}
\date{}
\date{}
\date{}
\date{today}
\date{}
\date{}
\date{}
\date{}
\date{}
\date{}
\date{}
\date{}
\date{today}
\date{}
\date{}
\date{}
\date{}
\date{}
\date{}
\date{}
\date{}
\date{today}
\date{}
\date{}
\date{}
\date{}
\date{}
\date{}
\date{}
\date{}
\date{today}
\date{}
\date{}
\date{}
\date{}
\date{}
\date{}
\date{}
\date{}
\date{today}
\date{}
\date{}
\date{}
\date{}
\date{}
\date{}
\date{}
\date{}
\date{today}
\date{}
\date{}
\date{}
\date{}
\date{}
\date{}
\date{}
\date{}
\date{today}
\date{}
\date{}
\date{}
\date{}
\date{}
\date{}
\date{}
\date{}
\date{today}
\date{}
\date{}
\date{}
\date{}
\date{}
\date{}
\date{}
\date{}
\date{today}
\date{}
\date{}
\date{}
\date{}
\date{}
\date{}
\date{}
\date{}
\date{today}
\date{}
\date{}
\date{}
\date{}
\date{}
\date{}
\date{}
\date{}
\date{today}
\date{}
\date{}
\date{}
\date{}
\date{}
\date{}
\date{}
\date{}
\date{today}
\date{}
\date{}
\date{}
\date{}
\date{}
\date{}
\date{}
\date{}
\date{today}
\date{}
\date{}
\date{}
\date{}
\date{}
\date{}
\date{}
\date{}
\date{today}
\date{}
\date{}
\date{}
\date{}
\date{}
\date{}
\date{}
\date{}
\date{today}
\date{}
\date{}
\date{}
\date{}
\date{}
\date{}
\date{}
\date{}
\date{today}
\date{}
\date{}
\date{}
\date{}
\date{}
\date{}
\date{}
\date{}
\date{today}
\date{}
\date{}
\date{}
\date{}
\date{}
\date{}
\date{}
\date{}
\date{today}
\date{}
\date{}
\date{}
\date{}
\date{}
\date{}
\date{}
\date{}
\date{today}
\date{}
\date{}
\date{}
\date{}
\date{}
\date{}
\date{}
\date{}
\date{today}
\date{}
\date{}
\date{}
\date{}
\date{}
\date{}
\date{}
\date{}
\date{today}
\date{}
\date{}
\date{}
\date{}
\date{}
\date{}
\date{}
\date{}
\date{today}
\date{}
\date{}
\date{}
\date{}
\date{}
\date{}
\date{}
\date{}
\date{today}
\date{}
\date{}
\date{}
\date{}
\date{}
\date{}
\date{}
\date{}
\date{today}
\date{}
\date{}
\date{}
\date{}
\date{}
\date{}
\date{}
\date{}
\date{today}
\date{}
\date{}
\date{}
\date{}
\date{}
\date{}
\date{}
\date{}
\date{today}
\date{}
\date{}
\date{}
\date{}
\date{}
\date{}
\date{}
\date{}
\date{today}
\date{}
\date{}
\date{}
\date{}
\date{}
\date{}
\date{}
\date{}
\date{today}
\date{}
\date{}
\date{}
\date{}
\date{}
\date{}
\date{}
\date{}
\date{today}
\date{}
\date{}
\date{}
\date{}
\date{}
\date{}
\date{}
\date{}
\date{today}
\date{}
\date{}
\date{}
\date{}
\date{}
\date{}
\date{}
\date{}
\date{today}
\date{}
\date{}
\date{}
\date{}
\date{}
\date{}
\date{}
\date{}
\date{today}
\date{}
\date{}
\date{}
\date{}
\date{}
\date{}
\date{}
\date{}
\date{today}
\date{}
\date{}
\date{}
\date{}
\date{}
\date{}
\date{}
\date{}
\date{today}
\date{}
\date{}
\date{}
\date{}
\date{}
\date{}
\date{}
\date{}
\date{today}
\date{}
\date{}
\date{}
\date{}
\date{}
\date{}
\date{}
\date{}
\date{today}
\date{}
\date{}
\date{}
\date{}
\date{}
\date{}
\date{}
\date{}
\date{today}
\date{}
\date{}
\date{}
\date{}
\date{}
\date{}
\date{}
\date{}
\date{today}
\date{}
\date{}
\date{}
\date{}
\date{}
\date{}
\date{}
\date{}
\date{today}
\date{}
\date{}
\date{}
\date{}
\date{}
\date{}
\date{}
\date{}
\date{today}
\date{}
\date{}
\date{}
\date{}
\date{}
\date{}
\date{}
\date{}
\date{today}
\date{}
\date{}
\date{}
\date{}
\date{}
\date{}
\date{}
\date{}
\date{today}
\date{}
\date{}
\date{}
\date{}
\date{}
\date{}
\date{}
\date{}
\date{today}
\date{}
\date{}
\date{}
\date{}
\date{}
\date{}
\date{}
\date{}
\date{today}
\date{}
\date{}
\date{}
\date{}
\date{}
\date{}
\date{}
\date{}
\date{today}
\date{}
\date{}
\date{}
\date{}
\date{}
\date{}
\date{}
\date{}
\date{today}
\date{}
\date{}
\date{}
\date{}
\date{}
\date{}
\date{}
\date{}
\date{today}
\date{}
\date{}
\date{}
\date{}
\date{}
\date{}
\date{}
\date{}
\date{today}
\date{}
\date{}
\date{}
\date{}
\date{}
\date{}
\date{}
\date{}
\date{today}
\date{}
\date{}
\date{}
\date{}
\date{}
\date{}
\date{}
\date{}
\date{today}
\date{}
\date{}
\date{}
\date{}
\date{}
\date{}
\date{}
\date{}
\date{today}
\date{}
\date{}
\date{}
\date{}
\date{}
\date{}
\date{}
\date{}
\date{today}
\date{}
\date{}
\date{}
\date{}
\date{}
\date{}
\date{}
\date{}
\date{today}
\date{}
\date{}
\date{}
\date{}
\date{}
\date{}
\date{}
\date{}
\date{today}
\date{}
\date{}
\date{}
\date{}
\date{}
\date{}
\date{}
\date{}
\date{today}
\date{}
\date{}
\date{}
\date{}
\date{}
\date{}
\date{}
\date{}
\date{today}
\date{}
\date{}
\date{}
\date{}
\date{}
\date{}
\date{}
\date{}
\date{today}
\date{}
\date{}
\date{}
\date{}
\date{}
\date{}
\date{}
\date{}
\date{today}
\date{}
\date{}
\date{}
\date{}
\date{}
\date{}
\date{}
\date{}
\date{today}
\date{}
\date{}
\date{}
\date{}
\date{}
\date{}
\date{}
\date{}
\date{today}
\date{}
\date{}
\date{}
\date{}
\date{}
\date{}
\date{}
\date{}
\date{today}
\date{}
\date{}
\date{}
\date{}
\date{}
\date{}
\date{}
\date{}
\date{today}
\date{}
\date{}
\date{}
\date{}
\date{}
\date{}
\date{}
\date{}
\date{today}
\date{}
\date{}
\date{}
\date{}
\date{}
\date{}
\date{}
\date{}
\date{today}
\date{}
\date{}
\date{}
\date{}
\date{}
\date{}
\date{}
\date{}
\date{today}
\date{}
\date{}
\date{}
\date{}
\date{}
\date{}
\date{}
\date{}
\date{today}
\date{}
\date{}
\date{}
\date{}
\date{}
\date{}
\date{}
\date{}
\date{today}
\date{}
\date{}
\date{}
\date{}
\date{}
\date{}
\date{}
\date{}
\date{today}
\date{}
\date{}
\date{}
\date{}
\date{}
\date{}
\date{}
\date{}
\date{today}
\date{}
\date{}
\date{}
\date{}
\date{}
\date{}
\date{}
\date{}
\date{today}
\date{}
\date{}
\date{}
\date{}
\date{}
\date{}
\date{}
\date{}
\date{today}
\date{}
\date{}
\date{}
\date{}
\date{}
\date{}
\date{}
\date{}
\date{today}
\date{}
\date{}
\date{}
\date{}
\date{}
\date{}
\date{}
\date{}
\date{today}
\date{}
\date{}
\date{}
\date{}
\date{}
\date{}
\date{}
\date{}
\date{today}
\date{}
\date{}
\date{}
\date{}
\date{}
\date{}
\date{}
\date{}
\date{today}
\date{}
\date{}
\date{}
\date{}
\date{}
\date{}
\date{}
\date{}
\date{today}
\date{}
\date{}
\date{}
\date{}
\date{}
\date{}
\date{}
\date{}
\date{today}
\date{}
\date{}
\date{}
\date{}
\date{}
\date{}
\date{}
\date{}
\date{today}
\date{}
\date{}
\date{}
\date{}
\date{}
\date{}
\date{}
\date{}
\date{today}
\date{}
\date{}
\date{}
\date{}
\date{}
\date{}
\date{}
\date{}
\date{today}
\date{}
\date{}
\date{}
\date{}
\date{}
\date{}
\date{}
\date{}
\date{today}
\date{}
\date{}
\date{}
\date{}
\date{}
\date{}
\date{}
\date{}
\date{today}
\date{}
\date{}
\date{}
\date{}
\date{}
\date{}
\date{}
\date{}
\date{today}
\date{}
\date{}
\date{}
\date{}
\date{}
\date{}
\date{}
\date{}
\date{today}
\date{}
\date{}
\date{}
\date{}
\date{}
\date{}
\date{}
\date{}
\date{today}
\date{}
\date{}
\date{}
\date{}
\date{}
\date{}
\date{}
\date{}
\date{today}
\date{}
\date{}
\date{}
\date{}
\date{}
\date{}
\date{}
\date{}
\date{today}
\date{}
\date{}
\date{}
\date{}
\date{}
\date{}
\date{}
\date{}
\date{today}
\date{}
\date{}
\date{}
\date{}
\date{}
\date{}
\date{}
\date{}
\date{today}
\date{}
\date{}
\date{}
\date{}
\date{}
\date{}
\date{}
\date{}
\date{today}
\date{}
\date{}
\date{}
\date{}
\date{}
\date{}
\date{}
\date{}
\date{today}
\date{}
\date{}
\date{}
\date{}
\date{}
\date{}
\date{}
\date{}
\date{today}
\date{}
\date{}
\date{}
\date{}
\date{}
\date{}
\date{}
\date{}
\date{today}
\date{}
\date{}
\date{}
\date{}
\date{}
\date{}
\date{}
\date{}
\date{today}
\date{}
\date{}
\date{}
\date{}
\date{}
\date{}
\date{}
\date{}
\date{today}
\date{}
\date{}
\date{}
\date{}
\date{}
\date{}
\date{}
\date{}
\date{today}
\date{}
\date{}
\date{}
\date{}
\date{}
\date{}
\date{}
\date{}
\date{today}
\date{}
\date{}
\date{}
\date{}
\date{}
\date{}
\date{}
\date{}
\date{today}
\date{}
\date{}
\date{}
\date{}
\date{}
\date{}
\date{}
\date{}
\date{today}
\date{}
\date{}
\date{}
\date{}
\date{}
\date{}
\date{}
\date{}
\date{today}
\date{}
\date{}
\date{}
\date{}
\date{}
\date{}
\date{}
\date{}
\date{today}
\date{}
\date{}
\date{}
\date{}
\date{}
\date{}
\date{}
\date{}
\date{today}
\date{}
\date{}
\date{}
\date{}
\date{}
\date{}
\date{}
\date{}
\date{today}
\date{}
\date{}
\date{}
\date{}
\date{}
\date{}
\date{}
\date{}
\date{today}
\date{}
\date{}
\date{}
\date{}
\date{}
\date{}
\date{}
\date{}
\date{today}
\date{}
\date{}
\date{}
\date{}
\date{}
\date{}
\date{}
\date{}
\date{today}
\date{}
\date{}
\date{}
\date{}
\date{}
\date{}
\date{}
\date{}
\date{today}
\date{}
\date{}
\date{}
\date{}
\date{}
\date{}
\date{}
\date{}
\date{today}
\date{}
\date{}
\date{}
\date{}
\date{}
\date{}
\date{}
\date{}
\date{today}
\date{}
\date{}
\date{}
\date{}
\date{}
\date{}
\date{}
\date{}
\date{today}
\date{}
\date{}
\date{}
\date{}
\date{}
\date{}
\date{}
\date{}
\date{today}
\date{}
\date{}
\date{}
\date{}
\date{}
\date{}
\date{}
\date{}
\date{today}
\date{}
\date{}
\date{}
\date{}
\date{}
\date{}
\date{}
\date{}
\date{today}
\date{}
\date{}
\date{}
\date{}
\date{}
\date{}
\date{}
\date{}
\date{today}
\date{}
\date{}
\date{}
\date{}
\date{}
\date{}
\date{}
\date{}
\date{today}
\date{}
\date{}
\date{}
\date{}
\date{}
\date{}
\date{}
\date{}
\date{today}
\date{}
\date{}
\date{}
\date{}
\date{}
\date{}
\date{}
\date{}
\date{today}
\date{}
\date{}
\date{}
\date{}
\date{}
\date{}
\date{}
\date{}
\date{today}
\date{}
\date{}
\date{}
\date{}
\date{}
\date{}
\date{}
\date{}
\date{today}
\date{}
\date{}
\date{}
\date{}
\date{}
\date{}
\date{}
\date{}
\date{today}
\date{}
\date{}
\date{}
\date{}
\date{}
\date{}
\date{}
\date{}
\date{today}
\date{}
\date{}
\date{}
\date{}
\date{}
\date{}
\date{}
\date{}
\date{today}
\date{}
\date{}
\date{}
\date{}
\date{}
\date{}
\date{}
\date{}
\date{today}
\date{}
\date{}
\date{}
\date{}
\date{}
\date{}
\date{}
\date{}
\date{today}
\date{}
\date{}
\date{}
\date{}
\date{}
\date{}
\date{}
\date{}
\date{today}
\date{}
\date{}
\date{}
\date{}
\date{}
\date{}
\date{}
\date{}
\date{today}
\date{}
\date{}
\date{}
\date{}
\date{}
\date{}
\date{}
\date{}
\date{today}
\date{}
\date{}
\date{}
\date{}
\date{}
\date{}
\date{}
\date{}
\date{today}
\date{}
\date{}
\date{}
\date{}
\date{}
\date{}
\date{}
\date{}
\date{today}
\date{}
\date{}
\date{}
\date{}
\date{}
\date{}
\date{}
\date{}
\date{today}
\date{}
\date{}
\date{}
\date{}
\date{}
\date{}
\date{}
\date{}
\date{today}
\date{}
\date{}
\date{}
\date{}
\date{}
\date{}
\date{}
\date{}
\date{today}
\date{}
\date{}
\date{}
\date{}
\date{}
\date{}
\date{}
\date{}
\date{today}
\date{}
\date{}
\date{}
\date{}
\date{}
\date{}
\date{}
\date{}
\date{today}
\date{}
\date{}
\date{}
\date{}
\date{}
\date{}
\date{}
\date{}
\date{today}
\date{}
\date{}
\date{}
\date{}
\date{}
\date{}
\date{}
\date{}
\date{today}
\date{}
\date{}
\date{}
\date{}
\date{}
\date{}
\date{}
\date{}
\date{today}
\date{}
\date{}
\date{}
\date{}
\date{}
\date{}
\date{}
\date{}
\date{today}
\date{}
\date{}
\date{}
\date{}
\date{}
\date{}
\date{}
\date{}
\date{today}
\date{}
\date{}
\date{}
\date{}
\date{}
\date{}
\date{}
\date{}
\date{today}
\date{}
\date{}
\date{}
\date{}
\date{}
\date{}
\date{}
\date{}
\date{today}
\date{}
\date{}
\date{}
\date{}
\date{}
\date{}
\date{}
\date{}
\date{today}
\date{}
\date{}
\date{}
\date{}
\date{}
\date{}
\date{}
\date{}
\date{today}
\date{}
\date{}
\date{}
\date{}
\date{}
\date{}
\date{}
\date{}
\date{today}
\date{}
\date{}
\date{}
\date{}
\date{}
\date{}
\date{}
\date{}
\date{today}
\date{}
\date{}
\date{}
\date{}
\date{}
\date{}
\date{}
\date{}
\date{today}
\date{}
\date{}
\date{}
\date{}
\date{}
\date{}
\date{}
\date{}
\date{today}
\date{}
\date{}
\date{}
\date{}
\date{}
\date{}
\date{}
\date{}
\date{today}
\date{}
\date{}
\date{}
\date{}
\date{}
\date{}
\date{}
\date{}
\date{today}
\date{}
\date{}
\date{}
\date{}
\date{}
\date{}
\date{}
\date{}
\date{today}
\date{}
\date{}
\date{}
\date{}
\date{}
\date{}
\date{}
\date{}
\date{today}
\date{}
\date{}
\date{}
\date{}
\date{}
\date{}
\date{}
\date{}
\date{today}
\date{}
\date{}
\date{}
\date{}
\date{}
\date{}
\date{}
\date{}
\date{today}
\date{}
\date{}
\date{}
\date{}
\date{}
\date{}
\date{}
\date{}
\date{today}
\date{}
\date{}
\date{}
\date{}
\date{}
\date{}
\date{}
\date{}
\date{today}
\date{}
\date{}
\date{}
\date{}
\date{}
\date{}
\date{}
\date{}
\date{today}
\date{}
\date{}
\date{}
\date{}
\date{}
\date{}
\date{}
\date{}
\date{today}
\date{}
\date{}
\date{}
\date{}
\date{}
\date{}
\date{}
\date{}
\date{today}
\date{}
\date{}
\date{}
\date{}
\date{}
\date{}
\date{}
\date{}
\date{today}
\date{}
\date{}
\date{}
\date{}
\date{}
\date{}
\date{}
\date{}
\date{today}
\date{}
\date{}
\date{}
\date{}
\date{}
\date{}
\date{}
\date{}
\date{today}
\date{}
\date{}
\date{}
\date{}
\date{}
\date{}
\date{}
\date{}
\date{today}
\date{}
\date{}
\date{}
\date{}
\date{}
\date{}
\date{}
\date{}
\date{today}
\date{}
\date{}
\date{}
\date{}
\date{}
\date{}
\date{}
\date{}
\date{today}
\date{}
\date{}
\date{}
\date{}
\date{}
\date{}
\date{}
\date{}
\date{today}
\date{}
\date{}
\date{}
\date{}
\date{}
\date{}
\date{}
\date{}
\date{today}
\date{}
\date{}
\date{}
\date{}
\date{}
\date{}
\date{}
\date{}
\date{today}
\date{}
\date{}
\date{}
\date{}
\date{}
\date{}
\date{}
\date{}
\date{today}
\date{}
\date{}
\date{}
\date{}
\date{}
\date{}
\date{}
\date{}
\date{today}
\date{}
\date{}
\date{}
\date{}
\date{}
\date{}
\date{}
\date{}
\date{today}
\date{}
\date{}
\date{}
\date{}
\date{}
\date{}
\date{}
\date{}
\date{today}
\date{}
\date{}
\date{}
\date{}
\date{}
\date{}
\date{}
\date{}
\date{today}
\date{}
\date{}
\date{}
\date{}
\date{}
\date{}
\date{}
\date{}
\date{today}
\date{}
\date{}
\date{}
\date{}
\date{}
\date{}
\date{}
\date{}
\date{today}
\date{}
\date{}
\date{}
\date{}
\date{}
\date{}
\date{}
\date{}
\date{today}
\date{}
\date{}
\date{}
\date{}
\date{}
\date{}
\date{}
\date{}
\date{today}
\date{}
\date{}
\date{}
\date{}
\date{}
\date{}
\date{}
\date{}
\date{today}
\date{}
\date{}
\date{}
\date{}
\date{}
\date{}
\date{}
\date{}
\date{today}
\date{}
\date{}
\date{}
\date{}
\date{}
\date{}
\date{}
\date{}
\date{today}
\date{}
\date{}
\date{}
\date{}
\date{}
\date{}
\date{}
\date{}
\date{today}
\date{}
\date{}
\date{}
\date{}
\date{}
\date{}
\date{}
\date{}
\date{today}
\date{}
\date{}
\date{}
\date{}
\date{}
\date{}
\date{}
\date{}
\date{today}
\date{}
\date{}
\date{}
\date{}
\date{}
\date{}
\date{}
\date{}
\date{today}
\date{}
\date{}
\date{}
\date{}
\date{}
\date{}
\date{}
\date{}
\date{today}
\date{}
\date{}
\date{}
\date{}
\date{}
\date{}
\date{}
\date{}
\date{today}
\date{}
\date{}
\date{}
\date{}
\date{}
\date{}
\date{}
\date{}
\date{today}
\date{}
\date{}
\date{}
\date{}
\date{}
\date{}
\date{}
\date{}
\date{today}
\date{}
\date{}
\date{}
\date{}
\date{}
\date{}
\date{}
\date{}
\date{today}
\date{}
\date{}
\date{}
\date{}
\date{}
\date{}
\date{}
\date{}
\date{today}
\date{}
\date{}
\date{}
\date{}
\date{}
\date{}
\date{}
\date{}
\date{today}
\date{}
\date{}
\date{}
\date{}
\date{}
\date{}
\date{}
\date{}
\date{today}
\date{}
\date{}
\date{}
\date{}
\date{}
\date{}
\date{}
\date{}
\date{today}
\date{}
\date{}
\date{}
\date{}
\date{}
\date{}
\date{}
\date{}
\date{today}
\date{}
\date{}
\date{}
\date{}
\date{}
\date{}
\date{}
\date{}
\date{today}
\date{}
\date{}
\date{}
\date{}
\date{}
\date{}
\date{}
\date{}
\date{today}
\date{}
\date{}
\date{}
\date{}
\date{}
\date{}
\date{}
\date{}
\date{today}
\date{}
\date{}
\date{}
\date{}
\date{}
\date{}
\date{}
\date{}
\date{today}
\date{}
\date{}
\date{}
\date{}
\date{}
\date{}
\date{}
\date{}
\date{today}
\date{}
\date{}
\date{}
\date{}
\date{}
\date{}
\date{}
\date{}
\date{today}
\date{}
\date{}
\date{}
\date{}
\date{}
\date{}
\date{}
\date{}
\date{today}
\date{}
\date{}
\date{}
\date{}
\date{}
\date{}
\date{}
\date{}
\date{today}
\date{}
\date{}
\date{}
\date{}
\date{}
\date{}
\date{}
\date{}
\date{today}
\date{}
\date{}
\date{}
\date{}
\date{}
\date{}
\date{}
\date{}
\date{today}
\date{}
\date{}
\date{}
\date{}
\date{}
\date{}
\date{}
\date{}
\date{today}
\date{}
\date{}
\date{}
\date{}
\date{}
\date{}
\date{}
\date{}
\date{today}
\date{}
\date{}
\date{}
\date{}
\date{}
\date{}
\date{}
\date{}
\date{today}
\date{}
\date{}
\date{}
\date{}
\date{}
\date{}
\date{}
\date{}
\date{today}
\date{}
\date{}
\date{}
\date{}
\date{}
\date{}
\date{}
\date{}
\date{today}
\date{}
\date{}
\date{}
\date{}
\date{}
\date{}
\date{}
\date{}
\date{today}
\date{}
\date{}
\date{}
\date{}
\date{}
\date{}
\date{}
\date{}
\date{today}
\date{}
\date{}
\date{}
\date{}
\date{}
\date{}
\date{}
\date{}
\date{today}
\date{}
\date{}
\date{}
\date{}
\date{}
\date{}
\date{}
\date{}
\date{today}
\date{}
\date{}
\date{}
\date{}
\date{}
\date{}
\date{}
\date{}
\date{today}
\date{}
\date{}
\date{}
\date{}
\date{}
\date{}
\date{}
\date{}
\date{today}
\date{}
\date{}
\date{}
\date{}
\date{}
\date{}
\date{}
\date{}
\date{today}
\date{}
\date{}
\date{}
\date{}
\date{}
\date{}
\date{}
\date{}
\date{today}
\date{}
\date{}
\date{}
\date{}
\date{}
\date{}
\date{}
\date{}
\date{today}
\date{}
\date{}
\date{}
\date{}
\date{}
\date{}
\date{}
\date{}
\date{today}
\date{}
\date{}
\date{}
\date{}
\date{}
\date{}
\date{}
\date{}
\date{today}
\date{}
\date{}
\date{}
\date{}
\date{}
\date{}
\date{}
\date{}
\date{today}
\date{}
\date{}
\date{}
\date{}
\date{}
\date{}
\date{}
\date{}
\date{today}
\date{}
\date{}
\date{}
\date{}
\date{}
\date{}
\date{}
\date{}
\date{today}
\date{}
\date{}
\date{}
\date{}
\date{}
\date{}
\date{}
\date{}
\date{today}
\date{}
\date{}
\date{}
\date{}
\date{}
\date{}
\date{}
\date{}
\date{today}
\date{}
\date{}
\date{}
\date{}
\date{}
\date{}
\date{}
\date{}
\date{today}
\date{}
\date{}
\date{}
\date{}
\date{}
\date{}
\date{}
\date{}
\date{today}
\date{}
\date{}
\date{}
\date{}
\date{}
\date{}
\date{}
\date{}
\date{today}
\date{}
\date{}
\date{}
\date{}
\date{}
\date{}
\date{}
\date{}
\date{today}
\date{}
\date{}
\date{}
\date{}
\date{}
\date{}
\date{}
\date{}
\date{today}
\date{}
\date{}
\date{}
\date{}
\date{}
\date{}
\date{}
\date{}
\date{today}
\date{}
\date{}
\date{}
\date{}
\date{}
\date{}
\date{}
\date{}
\date{today}
\date{}
\date{}
\date{}
\date{}
\date{}
\date{}
\date{}
\date{}
\date{today}
\date{}
\date{}
\date{}
\date{}
\date{}
\date{}
\date{}
\date{}
\date{today}
\date{}
\date{}
\date{}
\date{}
\date{}
\date{}
\date{}
\date{}
\date{today}
\date{}
\date{}
\date{}
\date{}
\date{}
\date{}
\date{}
\date{}
\date{today}
\date{}
\date{}
\date{}
\date{}
\date{}
\date{}
\date{}
\date{}
\date{today}
\date{}
\date{}
\date{}
\date{}
\date{}
\date{}
\date{}
\date{}
\date{today}
\date{}
\date{}
\date{}
\date{}
\date{}
\date{}
\date{}
\date{}
\date{today}
\date{}
\date{}
\date{}
\date{}
\date{}
\date{}
\date{}
\date{}
\date{today}
\date{}
\date{}
\date{}
\date{}
\date{}
\date{}
\date{}
\date{}
\date{today}
\date{}
\date{}
\date{}
\date{}
\date{}
\date{}
\date{}
\date{}
\date{today}
\date{}
\date{}
\date{}
\date{}
\date{}
\date{}
\date{}
\date{}
\date{today}
\date{}
\date{}
\date{}
\date{}
\date{}
\date{}
\date{}
\date{}
\date{today}
\date{}
\date{}
\date{}
\date{}
\date{}
\date{}
\date{}
\date{}
\date{today}
\date{}
\date{}
\date{}
\date{}
\date{}
\date{}
\date{}
\date{}
\date{today}
\date{}
\date{}
\date{}
\date{}
\date{}
\date{}
\date{}
\date{}
\date{today}
\date{}
\date{}
\date{}
\date{}
\date{}
\date{}
\date{}
\date{}
\date{today}
\date{}
\date{}
\date{}
\date{}
\date{}
\date{}
\date{}
\date{}
\date{today}
\date{}
\date{}
\date{}
\date{}
\date{}
\date{}
\date{}
\date{}
\date{today}
\date{}
\date{}
\date{}
\date{}
\date{}
\date{}
\date{}
\date{}
\date{today}
\date{}
\date{}
\date{}
\date{}
\date{}
\date{}
\date{}
\date{}
\date{today}
\date{}
\date{}
\date{}
\date{}
\date{}
\date{}
\date{}
\date{}
\date{today}
\date{}
\date{}
\date{}
\date{}
\date{}
\date{}
\date{}
\date{}
\date{today}
\date{}
\date{}
\date{}
\date{}
\date{}
\date{}
\date{}
\date{}
\date{today}
\date{}
\date{}
\date{}
\date{}
\date{}
\date{}
\date{}
\date{}
\date{today}
\date{}
\date{}
\date{}
\date{}
\date{}
\date{}
\date{}
\date{}
\date{today}
\date{}
\date{}
\date{}
\date{}
\date{}
\date{}
\date{}
\date{}
\date{today}
\date{}
\date{}
\date{}
\date{}
\date{}
\date{}
\date{}
\date{}
\date{today}
\date{}
\date{}
\date{}
\date{}
\date{}
\date{}
\date{}
\date{}
\date{today}
\date{}
\date{}
\date{}
\date{}
\date{}
\date{}
\date{}
\date{}
\date{today}
\date{}
\date{}
\date{}
\date{}
\date{}
\date{}
\date{}
\date{}
\date{today}
\date{}
\date{}
\date{}
\date{}
\date{}
\date{}
\date{}
\date{}
\date{today}
\date{}
\date{}
\date{}
\date{}
\date{}
\date{}
\date{}
\date{}
\date{today}
\date{}
\date{}
\date{}
\date{}
\date{}
\date{}
\date{}
\date{}
\date{today}
\date{}
\date{}
\date{}
\date{}
\date{}
\date{}
\date{}
\date{}
\date{today}
\date{}
\date{}
\date{}
\date{}
\date{}
\date{}
\date{}
\date{}
\date{today}
\date{}
\date{}
\date{}
\date{}
\date{}
\date{}
\date{}
\date{}
\date{today}
\date{}
\date{}
\date{}
\date{}
\date{}
\date{}
\date{}
\date{}
\date{today}
\date{}
\date{}
\date{}
\date{}
\date{}
\date{}
\date{}
\date{}
\date{today}
\date{}
\date{}
\date{}
\date{}
\date{}
\date{}
\date{}
\date{}
\date{today}
\date{}
\date{}
\date{}
\date{}
\date{}
\date{}
\date{}
\date{}
\date{today}
\date{}
\date{}
\date{}
\date{}
\date{}
\date{}
\date{}
\date{}
\date{today}
\date{}
\date{}
\date{}
\date{}
\date{}
\date{}
\date{}
\date{}
\date{today}
\date{}
\date{}
\date{}
\date{}
\date{}
\date{}
\date{}
\date{}
\date{today}
\date{}
\date{}
\date{}
\date{}
\date{}
\date{}
\date{}
\date{}
\date{today}
\date{}
\date{}
\date{}
\date{}
\date{}
\date{}
\date{}
\date{}
\date{today}
\date{}
\date{}
\date{}
\date{}
\date{}
\date{}
\date{}
\date{}
\date{today}
\date{}
\date{}
\date{}
\date{}
\date{}
\date{}
\date{}
\date{}
\date{today}
\date{}
\date{}
\date{}
\date{}
\date{}
\date{}
\date{}
\date{}
\date{today}
\date{}
\date{}
\date{}
\date{}
\date{}
\date{}
\date{}
\date{}
\date{today}
\date{}
\date{}
\date{}
\date{}
\date{}
\date{}
\date{}
\date{}
\date{today}
\date{}
\date{}
\date{}
\date{}
\date{}
\date{}
\date{}
\date{}
\date{today}
\date{}
\date{}
\date{}
\date{}
\date{}
\date{}
\date{}
\date{}
\date{today}
\date{}
\date{}
\date{}
\date{}
\date{}
\date{}
\date{}
\date{}
\date{today}
\date{}
\date{}
\date{}
\date{}
\date{}
\date{}
\date{}
\date{}
\date{today}
\date{}
\date{}
\date{}
\date{}
\date{}
\date{}
\date{}
\date{}
\date{today}
\date{}
\date{}
\date{}
\date{}
\date{}
\date{}
\date{}
\date{}
\date{today}
\date{}
\date{}
\date{}
\date{}
\date{}
\date{}
\date{}
\date{}
\date{today}
\date{}
\date{}
\date{}
\date{}
\date{}
\date{}
\date{}
\date{}
\date{today}
\date{}
\date{}
\date{}
\date{}
\date{}
\date{}
\date{}
\date{}
\date{today}
\date{}
\date{}
\date{}
\date{}
\date{}
\date{}
\date{}
\date{}
\date{today}
\date{}
\date{}
\date{}
\date{}
\date{}
\date{}
\date{}
\date{}
\date{today}
\date{}
\date{}
\date{}
\date{}
\date{}
\date{}
\date{}
\date{}
\date{today}
\date{}
\date{}
\date{}
\date{}
\date{}
\date{}
\date{}
\date{}
\date{today}
\date{}
\date{}
\date{}
\date{}
\date{}
\date{}
\date{}
\date{}
\date{today}
\date{}
\date{}
\date{}
\date{}
\date{}
\date{}
\date{}
\date{}
\date{today}
\date{}
\date{}
\date{}
\date{}
\date{}
\date{}
\date{}
\date{}
\date{today}
\date{}
\date{}
\date{}
\date{}
\date{}
\date{}
\date{}
\date{}
\date{today}
\date{}
\date{}
\date{}
\date{}
\date{}
\date{}
\date{}
\date{}
\date{today}
\date{}
\date{}
\date{}
\date{}
\date{}
\date{}
\date{}
\date{}
\date{today}
\date{}
\date{}
\date{}
\date{}
\date{}
\date{}
\date{}
\date{}
\date{today}
\date{}
\date{}
\date{}
\date{}
\date{}
\date{}
\date{}
\date{}
\date{today}
\date{}
\date{}
\date{}
\date{}
\date{}
\date{}
\date{}
\date{}
\date{today}
\date{}
\date{}
\date{}
\date{}
\date{}
\date{}
\date{}
\date{}
\date{today}
\date{}
\date{}
\date{}
\date{}
\date{}
\date{}
\date{}
\date{}
\date{today}
\date{}
\date{}
\date{}
\date{}
\date{}
\date{}
\date{}
\date{}
\date{today}
\date{}
\date{}
\date{}
\date{}
\date{}
\date{}
\date{}
\date{}
\date{today}
\date{}
\date{}
\date{}
\date{}
\date{}
\date{}
\date{}
\date{}
\date{today}
\date{}
\date{}
\date{}
\date{}
\date{}
\date{}
\date{}
\date{}
\date{today}
\date{}
\date{}
\date{}
\date{}
\date{}
\date{}
\date{}
\date{}
\date{today}
\date{}
\date{}
\date{}
\date{}
\date{}
\date{}
\date{}
\date{}
\date{today}
\date{}
\date{}
\date{}
\date{}
\date{}
\date{}
\date{}
\date{}
\date{today}
\date{}
\date{}
\date{}
\date{}
\date{}
\date{}
\date{}
\date{}
\date{today}
\date{}
\date{}
\date{}
\date{}
\date{}
\date{}
\date{}
\date{}
\date{today}
\date{}
\date{}
\date{}
\date{}
\date{}
\date{}
\date{}
\date{}
\date{today}
\date{}
\date{}
\date{}
\date{}
\date{}
\date{}
\date{}
\date{}
\date{today}
\date{}
\date{}
\date{}
\date{}
\date{}
\date{}
\date{}
\date{}
\date{today}
\date{}
\date{}
\date{}
\date{}
\date{}
\date{}
\date{}
\date{}
\date{today}
\date{}
\date{}
\date{}
\date{}
\date{}
\date{}
\date{}
\date{}
\date{today}
\date{}
\date{}
\date{}
\date{}
\date{}
\date{}
\date{}
\date{}
\date{today}
\date{}
\date{}
\date{}
\date{}
\date{}
\date{}
\date{}
\date{}
\date{today}
\date{}
\date{}
\date{}
\date{}
\date{}
\date{}
\date{}
\date{}
\date{today}
\date{}
\date{}
\date{}
\date{}
\date{}
\date{}
\date{}
\date{}
\date{today}
\date{}
\date{}
\date{}
\date{}
\date{}
\date{}
\date{}
\date{}
\date{today}
\date{}
\date{}
\date{}
\date{}
\date{}
\date{}
\date{}
\date{}
\date{today}
\date{}
\date{}
\date{}
\date{}
\date{}
\date{}
\date{}
\date{}
\date{today}
\date{}
\date{}
\date{}
\date{}
\date{}
\date{}
\date{}
\date{}
\date{today}
\date{}
\date{}
\date{}
\date{}
\date{}
\date{}
\date{}
\date{}
\date{today}
\date{}
\date{}
\date{}
\date{}
\date{}
\date{}
\date{}
\date{}
\date{today}
\date{}
\date{}
\date{}
\date{}
\date{}
\date{}
\date{}
\date{}
\date{today}
\date{}
\date{}
\date{}
\date{}
\date{}
\date{}
\date{}
\date{}
\date{today}
\date{}
\date{}
\date{}
\date{}
\date{}
\date{}
\date{}
\date{}
\date{today}
\date{}
\date{}
\date{}
\date{}
\date{}
\date{}
\date{}
\date{}
\date{today}
\date{}
\date{}
\date{}
\date{}
\date{}
\date{}
\date{}
\date{}
\date{today}
\date{}
\date{}
\date{}
\date{}
\date{}
\date{}
\date{}
\date{}
\date{today}
\date{}
\date{}
\date{}
\date{}
\date{}
\date{}
\date{}
\date{}
\date{today}
\date{}
\date{}
\date{}
\date{}
\date{}
\date{}
\date{}
\date{}
\date{today}
\date{}
\date{}
\date{}
\date{}
\date{}
\date{}
\date{}
\date{}
\date{today}
\date{}
\date{}
\date{}
\date{}
\date{}
\date{}
\date{}
\date{}
\date{today}
\date{}
\date{}
\date{}
\date{}
\date{}
\date{}
\date{}
\date{}
\date{today}
\date{}
\date{}
\date{}
\date{}
\date{}
\date{}
\date{}
\date{}
\date{today}
\date{}
\date{}
\date{}
\date{}
\date{}
\date{}
\date{}
\date{}
\date{today}
\date{}
\date{}
\date{}
\date{}
\date{}
\date{}
\date{}
\date{}
\date{today}
\date{}
\date{}
\date{}
\date{}
\date{}
\date{}
\date{}
\date{}
\date{today}
\date{}
\date{}
\date{}
\date{}
\date{}
\date{}
\date{}
\date{}
\date{today}
\date{}
\date{}
\date{}
\date{}
\date{}
\date{}
\date{}
\date{}
\date{today}
\date{}
\date{}
\date{}
\date{}
\date{}
\date{}
\date{}
\date{}
\date{today}
\date{}
\date{}
\date{}
\date{}
\date{}
\date{}
\date{}
\date{}
\date{today}
\date{}
\date{}
\date{}
\date{}
\date{}
\date{}
\date{}
\date{}
\date{today}
\date{}
\date{}
\date{}
\date{}
\date{}
\date{}
\date{}
\date{}
\date{today}
\date{}
\date{}
\date{}
\date{}
\date{}
\date{}
\date{}
\date{}
\date{today}
\date{}
\date{}
\date{}
\date{}
\date{}
\date{}
\date{}
\date{}
\date{today}
\date{}
\date{}
\date{}
\date{}
\date{}
\date{}
\date{}
\date{}
\date{today}
\date{}
\date{}
\date{}
\date{}
\date{}
\date{}
\date{}
\date{}
\date{today}
\date{}
\date{}
\date{}
\date{}
\date{}
\date{}
\date{}
\date{}
\maketitle

\section{Introduction}

The Coupled-Channels (CC) method is the most powerful tool to study nuclear
reactions. However, it becomes extremely complicated when it is necessary to
include a large number of channels in the calculation. In situations where one
is only interested in a detailed description of a single channel, e.g. in
elastic scattering, one can resort to the polarization potential approach. It
consists of deriving a potential, to be added to the Hamiltonian of the
elastic channel which leads to the same elastic wave function as that obtained
by solution of the CC equations. The natural framework to derive this
potential is Feschbach formalism \cite{Fes62}. Although the exact derivation
of the polarization potential may be as difficult as solving the CC problem,
in some situations it is possible to find good approximations for it.

Approximate polarization potentials have been derived by several authors, for
collisions at near-barrier energies \cite{Vpol-LE, AGN94, CDH95}\ and energies
well above the barrier \cite{Ama85}. A serious drawback of these potentials is
that they are non-local and \textit{l}- and \textit{E}-dependent. Although any
non-local potential can be replaced by a trivially equivalent local one, the
latter present poles and has an artificial dependence on the quantum numbers
of the elastic wave function.

For practical purposes, it is convenient to have local polarization potentials
independent of $l$, in order that it could be used in standard computer codes.
Several approaches have been proposed to achieve this goal. In a recent paper,
Lubian and Nunes \cite{LuN07}\ tested the validity of the approach of Thompson
\textit{et al}. \cite{TNL89}, in the case of breakup coupling, which is very
important in collisions of weakly bound projectiles \cite{CGH06}. In the
present paper we extend this study to other procedures to derive
\textit{l}-independent potentials in the case of $^{11}\mathrm{Li}%
+^{12}\mathrm{C}$ scattering at near-barrier energies. For simplicity, we
represent the continuum by an effective bound channel, restricting ourselves
to two-channels. Although this is not an appropriate description of the
continuum, it is suitable for the qualitative purposes of the present work
since this effective state also leads to a long range polarization potential.

This paper is organized as follows. In sect. 2 we give a brief description of
the polarization potential, according to Feshbach's formalism. In sect. 3 we
discuss different prescriptions suggested to derive \textit{l}-independent
local potentials without poles. In sect. 4 we apply these prescriptions to
$^{11}\mathrm{Li}+^{12}\mathrm{C}$ scattering and compare the results with the
corresponding ones obtained with the CC method. Finally, in sect. 5 we
summarize the conclusions of this work.

\section{Feshbach's formalism for Polarization potentials}

Let us consider a collision described in terms of the collision degrees of
freedom (projectile-target separation vector), $\mathbf{r},$ and a set of
intrinsic degrees of freedom, represented by $\xi.$ The scattering wave
function $\Psi^{(+)}(\mathbf{r},\xi)$\ satisfies a Schr\"{o}dinger equation
with the total Hamiltonian%
\[
\mathbb{H}=H+h+\mathbb{V}.
\]
Above, $H$ depends only on the collision degrees of freedom$,~h$ acts only on
the intrinsic space and $\mathbb{V}$\ couples collision with intrinsic degrees
of freedom. It is convenient to perform the channel expansion of the
scattering state, in terms of the eigenstates of $h$, which satisfy the
equation%
\[
h\left\vert n\right)  =\varepsilon_{n}\left\vert n\right)  ,
\]
in the form
\begin{equation}
\left\vert \Psi^{(+)}\right\rangle =\sum_{\alpha=0}^{n}\left\vert \psi
_{\alpha}^{(+)}\right\rangle ~\left\vert \alpha\right)  . \label{ch-expan}%
\end{equation}
The intrinsic space can be divided in two complementary parts by the action of
the projectors%

\begin{equation}
P=|0)(0|;\;\;\;\;\;\;Q=\sum_{\alpha=1}^{n}|\alpha)(\alpha|. \label{PQ}%
\end{equation}
Above, $|0)$\ stands for the ground state of $h$ and we take $\varepsilon
_{0}=0.$\ These projectors have the properties
\begin{equation}
P^{2}=P,\;Q^{2}=Q;\;QP=PQ=0~\mathrm{and}\;\;P+Q=1~.\;~ \label{prop-proj-3}%
\end{equation}
Acting with these projectors on the scattering state, one gets%

\begin{align}
P~\left\vert \Psi^{(+)}\right\rangle  &  =\left\vert \psi_{0}^{(+)}%
\right\rangle ~\left\vert 0\right)  \equiv\left\vert \Psi_{P}\right\rangle
,\label{P-Psi}\\
Q~\left\vert \Psi^{(+)}\right\rangle  &  =\sum_{\alpha=1}^{n}\left\vert
\psi_{\alpha}^{(+)}\right\rangle ~\left\vert \alpha\right)  \equiv\left\vert
\Psi_{Q}\right\rangle ~,\label{Q-Psi}\\
\left\vert \Psi^{(+)}\right\rangle  &  =\left\vert \Psi_{P}\right\rangle
+\left\vert \Psi_{Q}\right\rangle .\label{Psi}%
\end{align}
Applying these projectors on the Schr\"{o}dinger equation, using
Eq.(\ref{Psi}) and rearranging the terms, one obtains
\begin{align}
\left[  E-H\right]  ~|\Psi_{P}\rangle &  =P\mathbb{V}Q|\Psi_{Q}\rangle
~.\label{proj1}\\
\left[  E-Q\mathbb{H}Q\right]  ~|\Psi_{Q}\rangle &  =Q\mathbb{V}P|\Psi
_{P}\rangle~.\label{proj2}%
\end{align}
Note that $PHQ=QHP=0$.

\bigskip Eq.( \ref{proj1}) can be used to define the polarization potential
operator viz,%
\begin{equation}
P\mathbb{V}Q|\Psi_{Q}\rangle=U^{pol}~|\Psi_{P}\rangle.\label{nova-eq}%
\end{equation}
This is the basis of the calculation of Refs. \cite{LuN07,CRH94} as they
employ a numerical solution of the CC equations and make use of the above
definition. To derive an explicit expression for the polarization potential
operator, one first derive an expression for the projected state $|\Psi
_{Q}\rangle$ by multiplying Eq.(\ref{proj2}) from the left with the Green's
function
\begin{equation}
\mathbb{G}_{QQ}^{(+)}=\frac{1}{E-Q\mathbb{H}Q+i\varepsilon}.\label{gr1}%
\end{equation}
The result is
\begin{equation}
|\Psi_{Q}\rangle=\mathbb{G}_{QQ}^{(+)}~Q\mathbb{V}P~|\Psi_{P}\rangle
~.\label{PsiQ}%
\end{equation}
Inserting the above equation in Eq.(\ref{proj1}), we get
\begin{equation}
\left[  E-\mathbb{H}_{eff}\right]  ~|\Psi_{P}\rangle=0.\label{Sch-Hef}%
\end{equation}
Above, $\mathbb{H}_{eff}$ is the effective Hamiltonian
\begin{equation}
\mathbb{H}_{eff}=PHP+P\mathbb{V}Q~\mathbb{G}_{QQ}^{(+)}~Q\mathbb{V}%
P~.\label{Hef}%
\end{equation}
Taking the scalar product of Eq.(\ref{Sch-Hef}) with $(0|$, using the explicit
form of $P$ (Eq.(\ref{PQ})) and replacing $(0|\Psi^{(+)}\rangle=|\psi
_{0}\rangle$, we obtain the Schr\"{o}dinger equation for the elastic wave
function in the space of the collision degree of freedom
\begin{equation}
(K+U^{opt}+U^{pol})~|\psi_{0}\rangle=E\,~|\psi_{0}\rangle.\label{schopol}%
\end{equation}
Above, $K$ is the kinetic energy operation, and the optical, $U^{opt}$, and
the polarization potentials, $U^{pol}$, are
\begin{align}
U^{opt} &  =\left(  0|H-K|0\right)  \label{vopt}\\
U^{pol} &  =(0|~\mathbb{V}Q~\mathbb{G}_{QQ}^{(+)}~Q~\mathbb{V}|0).\label{vpol}%
\end{align}
The above form for the polarization potential is numerically identical to the
one defined by Eq.(\ref{nova-eq})

As a consequence of the analytical structure of the Green operator (the
presence of $i\epsilon$ in Eq.( \ref{gr1})), we can immediately write:%
\begin{equation}
\operatorname{Im}\left\{  \mathbb{G}_{QQ}^{(+)}\right\}  =-\pi~\delta\left(
E-Q\mathbb{H}Q\right)  \label{ImG_QQ}%
\end{equation}
while the real part is%
\begin{equation}
\operatorname{Re}\left\{  \mathbb{G}_{QQ}^{(+)}\right\}  =\mathcal{P}~\left\{
\frac{1}{E-Q\mathbb{H}Q}\right\}  .\label{ReG_QQ}%
\end{equation}
This last equation can be rewritten as%
\[
\operatorname{Re}\left\{  \mathbb{G}_{QQ}^{(+)}\right\}  =-\mathcal{P}\int
dz~\frac{\delta\left(  z-Q\mathbb{H}Q\right)  }{z-E},
\]
which, with the help of Eq.(\ref{ImG_QQ}) gives the desired result%
\begin{equation}
\operatorname{Re}\left\{  \mathbb{G}_{QQ}^{(+)}(E)\right\}  =\frac{1}{\pi
}~\mathcal{P}\int dz~\frac{\operatorname{Im}\left\{  \mathbb{G}_{QQ}%
^{(+)}(z)\right\}  }{z-E}.\label{RelDisp1}%
\end{equation}
When Eqs.(\ref{ImG_QQ}) and (\ref{RelDisp1}) are used for the Green function
in Eq.(\ref{gr1}), we obtain the operator form of the dispersion relation,%

\begin{equation}
\operatorname{Re}\left\{  U^{pol}(E)\right\}  =\frac{1}{\pi}~\mathcal{P}\int
dz~\frac{\operatorname{Im}\left\{  U(z)\right\}  }{(z-E)}. \label{RelDisp2}%
\end{equation}
One would expect the calculated polarization potential to satisfy the above
relation. Since one usually resorts to several approximations to derive a
local \textit{l}-independent polarization potential, the above relation may
eventually be broken. However, it does supply an important check on numerical
evaluation of the polarization potential, just like unitarity and the S-matrix.

For practical purposes, it is convenient to write Eq.(\ref{schopol}) in the
coordinate representation. While the optical potential is generally taken to
be local and energy-dependent (owing to the effect of exchange non-locality,
removed to obtain a non-dispersive local equivalent potential \cite{CPH97}),
the non-locality of $\mathbb{G}_{QQ}^{(+)}$ leads to a non-local polarization
potential. One obtains the equation,
\begin{equation}
\left[  K+U^{opt}(\mathbf{r})\right]  ~\psi_{0}(\mathbf{r})+\int
U^{pol}(\mathbf{r},\mathbf{r}^{\prime})~\psi_{0}(\mathbf{r}^{\prime}%
)~d^{3}\mathbf{r}^{\prime}=E~\psi_{0}(\mathbf{r}),
\end{equation}
with
\begin{equation}
U^{pol}(\mathbf{r},\mathbf{r}^{\prime})=\left\langle \mathbf{r}\right\vert
~\left[  (0|~\mathbb{V}Q~\mathbb{G}_{QQ}^{(+)}~Q~\mathbb{V}|0)\right]
~\left\vert \mathbf{r}^{\prime}\right\rangle .\label{Upol-rr}%
\end{equation}

For some applications one replaces $U^{pol}(\mathbf{r},\mathbf{r}^{\prime})$
by the trivially equivalent local potential,
\begin{equation}
{\bar{U}}^{pol}(\mathbf{r})=\frac{1}{\psi_{0}(\mathbf{r})}\ \int
d^{3}\mathbf{r}^{\prime}\ U^{pol}(\mathbf{r},\mathbf{r}^{\prime})\ \psi
_{0}(\mathbf{r^{\prime}}). \label{loc-pol}%
\end{equation}
This potential has some undesirable features. Firstly, it has poles where
$\psi_{0}(\mathbf{r})$ vanishes. Secondly, the above procedure introduces
artificial dependences on the quantum numbers of $\psi_{0}.$\ These points
will be discussed in further details in the next section.

We should mention that our general expression for $U^{pol}(\mathbf{r}%
,\mathbf{r}^{\prime})$ (Eq. (\ref{Upol-rr})) can account for the most general
coupled channels situation. In particular, in the CDCC discussed in Ref.
\cite{LuN07}, the breakup continuum is discretized into an orthonormalized set
of bins, which would then span the \textit{Q}-space. The polarization
potential for this case, after writing for Q,%
\begin{equation}
Q=\sum_{b}~\left\vert b\right)  \left(  b\right\vert ,\label{Q-CDCC}%
\end{equation}
where $\left\vert b\right)  $ designates the $b^{th}$ bin, and assuming that
the coupling is local in the \textbf{r}-space, becomes%
\begin{equation}
U^{pol}(\mathbf{r},\mathbf{r}^{\prime})=\sum_{bb^{\prime}}~\mathcal{F}%
_{b}(\mathbf{r)}~~\mathbb{G}_{bb^{\prime}}^{(+)}(\mathbf{r},\mathbf{r}%
^{\prime})~\mathcal{F}_{b^{\prime}}(\mathbf{r}^{\prime}).\label{Vpol-CDCC1}%
\end{equation}
Above,%
\begin{equation}
\mathbb{G}_{bb^{\prime}}^{(+)}(\mathbf{r},\mathbf{r}^{\prime})=\left\langle
\mathbf{r}\right\vert ~\left[  (b|~\frac{1}{E-Q\mathbb{H}Q+i\varepsilon
}~|b^{\prime})\right]  ~\left\vert \mathbf{r}^{\prime}\right\rangle
\label{Gbb'}%
\end{equation}
and
\begin{equation}
\mathcal{F}_{b}(\mathbf{r)=}(0~|\mathbb{V}(\mathbf{r)}|b),\label{Fb}%
\end{equation}
with an analogous expression for $\mathcal{F}_{b^{\prime}}(\mathbf{r}^{\prime
}~).$ If we neglect continuum-continuum coupling and the width of the bins,
the Green's function becomes diagonal and Eq.(\ref{Vpol-CDCC1}) becomes,%
\begin{equation}
U^{pol}(\mathbf{r},\mathbf{r}^{\prime})=\sum_{b}~\mathcal{F}_{b}%
(\mathbf{r)}~~\mathbb{G}_{b}^{(+)}(\mathbf{r},\mathbf{r}^{\prime}%
)~\mathcal{F}_{b}(\mathbf{r}^{\prime}~).\label{Vpol-CDCC2}%
\end{equation}
Clearly both expressions for $U^{pol}$ above are highly non-local by
construction. In addition, the inclusion of continuum-continuum couplings
produces further non-local effects \cite{LuN07}. It is easier to deal with an
$U^{pol}$ having the form of Eq.(\ref{Vpol-CDCC2}). How to find an equivalent
no-continuum-continuum-coupling polarization potential? To answer this
question we rely on our recent work on the excitation of giant resonances in
heavy ion reactions. One usually excites a given state, which itself is
coupled to many other excited states. Using the exit doorway idea
\cite{CRH94}, namely the excitation of these other states from the ground
state proceeds from the exit doorway(s), one is bound to attach a width to the
exit doorways. Labelling the doorway states by $d$, the Green function takes
the diagonal form%
\begin{equation}
U^{pol}(\mathbf{r},\mathbf{r}^{\prime})=\sum_{d}~~\mathcal{F}_{d}%
(\mathbf{r)}~~\mathbb{G}_{d}^{(+)}(\mathbf{r},\mathbf{r}^{\prime}%
)~\mathcal{F}_{d}(\mathbf{r}^{\prime}~).\label{Vpol-CDCC3}%
\end{equation}
The energies
\[
\epsilon_{d}=\left(  d\right\vert h\left\vert d\right)  ,
\]
appearing in
\begin{equation}
\mathbb{G}_{d}^{(+)}=\frac{1}{E-\epsilon_{d}-~\left(  d\right\vert H\left\vert
d\right)  ~+i\varepsilon}\label{Gd}%
\end{equation}
are complex. The width of any given \textit{d}-state measures the strength of
the continuum-continuum coupling. The above expression for $U^{pol}%
(\mathbf{r},\mathbf{r}^{\prime})$ with complex $\epsilon_{d}$, should be a
faithful representation the full Green function WITH continuum-continuum
coupling. In the following we consider a much simpler two-channel case to
discuss the notion of equivalent \textit{l}-independent polarization
potential, and leave the discussion of the continuum-continuum case to a
future publication.

\bigskip One frequently perform angular momentum projections in the
Schr\"{o}dinger equation. For the simple case of a scalar Hamiltonian with
scalar coupling, one makes the expansions%
\begin{subequations}
\begin{align}
\psi_{0}^{(+)}(\mathbf{r)} &  \mathbf{=}\sum_{l}\frac{u_{l}(kr)}{r}%
~Y_{lm}(\mathbf{\hat{r}}),\label{upolbar-l}\\
U^{pol}(\mathbf{r},\mathbf{r}^{\prime}) &  =\frac{1}{rr^{\prime}}\sum
_{l}Y_{lm}(\mathbf{\hat{r}})~{U}_{l}^{pol}(r,r^{\prime})~Y_{lm}(\mathbf{\hat
{r}}^{\prime}).\label{upol-l}%
\end{align}
The angular-momentum projected version of Eq.(\ref{loc-pol}) is the local
equivalent polarization potential,
\end{subequations}
\begin{equation}
{U}_{l}^{pol}(r)=\frac{1}{u_{l}(kr)}\int dr^{\prime}~U_{l}^{pol}(r,r^{\prime
})~u_{l}(kr^{\prime}).\label{Upol-barl}%
\end{equation}
This potential should be included in the Hamiltonian for optical model
calculations of the elastic radial wave function, $u_{l}(r).$ It this way, one
needs a different polarization potential for each partial wave.

The polarization potential of Eq.(\ref{Upol-barl}) is not very useful, since
it requires the knowledge of the exact radial wave function in the elastic
channel, $u_{l}(kr).$ Determining $u_{l}(kr)$ is as hard as solving the
original CC equations.\ A rather widely used approximation for the
\textit{l}-dependent local polarization potential consists of replacing
$u_{l}(kr)$ by the optical radial wave function, $w_{l}(kr),$ which is the
solution of the partial-wave projected Schr\"{o}dinger equation with the
polarization potential switched off. This potential, denoted by ${\bar{U}}%
_{l}$, is \
\begin{equation}
{\bar{U}}_{l}(r)=\frac{1}{w_{l}(kr)}\int dr^{\prime}~U_{l}^{pol}(r,r^{\prime
})~w_{l}(kr^{\prime}).\label{TELLP}%
\end{equation}

A further inconvenience of trivially local equivalent potentials is that they
have poles wherever the radial wave function vanishes. In what follows we
discuss three different prescriptions for obtaining $l$- independent
polarization potentials free of poles$.$

\section{Local, \textit{l}-independent polarization potentials}

\subsection{The prescription of Thompson \textit{et al.}}

Thompson \textit{et al.} \cite{TNL89} proposed the following definition of an
$l$-independent version of the polarization potential,
\begin{equation}
U^{T}(r)=\frac{\sum_{l}~(2l+1)~T_{l}~|u_{l}(kr)|^{2}~U_{l}^{pol} (r)}{\sum
_{l}~(2l+1)~T_{l}~|u_{l}(kr)|^{2}~}, \label{VpolT}%
\end{equation}
where $T_{l}$ is the transmission coefficient in the elastic channel for the
$l^{th}$ partial-wave. The above definition of $U^{T}$ guarantees that no
poles remain in the polarization potential, which arise from the presence of
$u_{l}(r)$ in Eq. (\ref{VpolT}). Furthermore, owing to the presence of $T_{l}$
in the $l$-sum, only values of $l$ where $T_{l}$ is close to unity will
contribute. Of course the probability density $|u_{l}(r)|^{2}$ is small for
small values of $l$ due to absorption. Thus, in the prescription of Thompson
\textit{et al.}, Eq. (\ref{VpolT}) should contain contributions of $l$ in the
vicinity of the grazing one. We doubt that this is guaranteed always since
there is interference effects in $|u_{l}(r)|^{2}$ which may end up allowing
the contribution of small $l$ as well. This prescription has been recently
used in ref. \cite{LuN07} in the context of the Continuum Discretized Coupled
Channels calculation of break up and elastic scattering of $^{8}$B.

\subsection{A modified form of the prescription of Thompson \textit{et al.}}

We now introduce a slightly modified version of the above discussed
prescription, by substituting $T_{l}$ by its derivative with respect to $l$, namely,%

\begin{equation}
U^{MT}(r)=\frac{\sum_{l}~(2l+1)~\left(  dT_{l}/dl\right)  ~|u_{l}%
(kr)|^{2}~U_{l}^{pol}(r)}{\sum_{l}~(2l+1)~\left(  dT_{l}/dl\right)
~~|u_{l}(kr)|^{2}~}.\label{VpolTM}%
\end{equation}
The modified Thompson prescription (MT) above guarantees the contribution of
the $l-$values around the grazing one, regardless to the behavior
of\ $|u_{l}(r)|^{2}$. One physical motivation for choosing $dT_{l}/dl$ instead
of $T_{l}$ is that the DWBA amplitudes of non-elastic processes in the
adiabatic limit do behave as the $l-$derivative of the elastic $S$-matrix
elements, which enters in the definition of $T_{l}=1-|S_{l}|^{2}$.

\subsection{The semiclassical prescription}

This prescription relies on the semiclassical idea that the orbital angular
momentum, if treated classically, should be related to $r$ and the energy $E$
through the definition of the classical turning point $r_{t}$, namely%

\begin{equation}
\frac{\hbar^{2}}{2%
\mu
~r_{t}^{2}}~l(l+1)+U(r_{t})=E, \label{rt}%
\end{equation}
where $U$ is the real part of the optical potential containing the nuclear and
the Coulomb pieces. It is clear that, for a given collision energy,
$r_{t}\equiv r_{t}(l)$ is a function of $l.$~The prescription consists of
identifying $r\equiv r_{t}(l)$ in order to build the $l$-independent
polarization potential,%

\begin{equation}
U^{SC}(r)={\bar{U}}_{l}^{pol}(r_{t}(l)).\label{VpolSC}%
\end{equation}
Using Eq.(\ref{VpolSC}), one gets the potential $U^{SC}$ at a discrete set of
\textit{r}-values (one for each partial-wave) and interpolating between these
points one obtains a continuous function.

All three \textit{l}-independent potentials discussed above contain a further
energy dependence, besides the one that arises from the green function in the
polarization potential. This extra energy dependence is non-dispersive and
thus could render the applicability of the dispersion relation questionable.
We shall verify this point in the following section.

In all of the above prescriptions the starting point is the
trivially-equivalent local potential obtained from angular momentum projected
versions of Eqs.(\ref{Upol-rr}) and (\ref{loc-pol}), or the solution of the CC
equations used in Eq.(\ref{nova-eq}). This will be shown in detail in the next section.

\section{Application in the $^{11}\mathrm{Li+}^{12}\mathrm{C}$ scattering}

We have chosen the system $^{11}$Li + $^{12}$C to test the different local
approximations for the polarization potential discussed in the previous
sections. We have decided not to consider the coupling to the continuum
explicitly, but rather representing the continuum by a single bound effective
channel. It is clear that this is not an appropriate representation for the
continuum. One obvious shortcoming of this model is that it does not contain
continuum-continuum couplings, which is known to play an important role in
nuclear reactions with weakly bound projectiles \cite{CGH06}. The
justification for considering this schematic application is that our purpose
is to compare different approximations for the polarization potential, rather
than to perform quantitative calculations of elastic or fusion cross sections.

The two CC equations used in our calculation can be easily readout from Eqs.
(\ref{proj1}) and (\ref{proj2}). We denote the two intrinsic states by
$\left\vert 0\right)  $\ (ground state) and $\left\vert 1\right)  $ (excited
state). For simplicity, the numerical calculations are performed in the sudden
limit. In this limit, the excitation energy of the continuum states are
neglected ($\varepsilon_{0}=\varepsilon_{1}=0$). The projectors are
$P=\left\vert 0\right)  \left(  0\right\vert $ and $Q=\left\vert 1\right)
\left(  1\right\vert $ and , after taking the optical potentials in the two
channels to be the same, Eqs. (\ref{proj1}) and (\ref{proj2}) become
\begin{align}
\left[  E-K-U^{opt}(\mathbf{r})\right]  ~\Psi_{0}(\mathbf{r}) &
=\mathcal{F}(\mathbf{r})~\Psi_{1}(\mathbf{r})\label{proj3}\\
\left[  E-K-U^{opt}(\mathbf{r})\right]  ~\Psi_{1}(\mathbf{r}) &
=\mathcal{F}(\mathbf{r})~\Psi_{0}(\mathbf{r}).\label{proj4}%
\end{align}
Above, $\mathcal{F}(\mathbf{r})$ is the complex and symmetric form factor%
\begin{equation}
\mathcal{F}(\mathbf{r})=\int d^{3}\mathbf{x}~\varphi_{0}(\mathbf{x}%
)~\mathbb{V}(\mathbf{r},\mathbf{x})~\varphi_{1}(\mathbf{x}),\label{Form1}%
\end{equation}
evaluated with the coupling interaction%
\[
\mathbb{V}(\mathbf{r},\mathbf{x})=U_{f_{1}T}(\mathbf{r}_{f_{1}T})+U_{f_{2}%
T}(\mathbf{r}_{f_{2}T})-U^{opt}(\mathbf{r}).
\]
Since the intrinsic states are orthogonal and $U^{opt}(\mathbf{r})$\ does not
depend on \textbf{x}, Eq.(\ref{Form1}) reduces to%
\begin{equation}
\mathcal{F}(\mathbf{r})=\int d^{3}\mathbf{x}~\varphi_{0}(\mathbf{x})~\left[
U_{f_{1}T}(\mathbf{r}_{f_{1}T})+U_{f_{2}T}(\mathbf{r}_{f_{2}T})\right]
~\varphi_{1}(\mathbf{x}).\label{Form2}%
\end{equation}
For the present application, we assume that $^{11}$Li breaks up into two
fragments. The first, $f_{1},$ corresponds to a neutron pair, which we treat
as a single particle (di-neutron). The other, $f_{2},$ is the $^{9}%
$Li$\mathrm{-core.}$ In this way, the coordinates appearing in
Eqs.(\ref{Form1}) and (\ref{Form2}) are
\begin{align*}
\mathbf{r}_{f_{1}T} &  =\mathbf{r+\gamma}_{1}\mathbf{x,~~~~\gamma}_{1}%
=\frac{9}{11}\\
\mathbf{r}_{f_{2}T} &  =\mathbf{r+\gamma}_{2}\mathbf{x,~~~~\gamma}_{2}%
=-\frac{2}{11}.
\end{align*}
and the projectile-target separation vector, $\mathbf{r}$. $U_{f_{1}%
T}(\mathbf{r}_{f_{1}T}),U_{f_{2}T}(\mathbf{r}_{f_{2}T})$ and $U^{opt}%
(\mathbf{r})$ are the corresponding interactions. Of course $\mathcal{F}%
(\mathbf{r})$ can be evaluated exactly numerically once these potentials are
given and the single particle wave functions of the halo neutron in the ground
and in the excited state, $\varphi_{0}(\mathbf{x})=(\mathbf{x}\left\vert
0\right)  $ and $\varphi_{1}(\mathbf{x})=(\mathbf{x}\left\vert 1\right)  ,$
are used to evaluate the integral in Eq.(\ref{Form1}). For our purposes we
parametrize $\mathcal{F}(\mathbf{r})$ as%
\begin{equation}
\mathcal{F}(\mathbf{r})=\mathcal{F}_{0}~\exp\left[  -\frac{r}{\mathbf{\gamma
}_{1}\alpha}\right]  ,\label{ffact}%
\end{equation}
with
\begin{equation}
\alpha=\frac{\hbar}{\sqrt{2\mu_{1-2}B}}.\label{alpha}%
\end{equation}
Above, $\mu_{1-2}$ is the reduced mass of the fragments inside the projectile
and $B$ is the breakup threshold. For $^{11}$Li, $\mu_{1-2}=18~m_{0}/11,$
$B=0.376$ MeV, and one gets $\alpha=5.83$ fm. For the purpose of simplicity,
we take $\mathcal{F}_{0}$ to be real. We also ignore the Coulomb coupling
altogether. A qualitative justification for the approximation of
Eq.(\ref{ffact}) is given in the appendix.

\bigskip

\bigskip The optical potentials can be written as%
\[
U^{opt}(\mathbf{r})=U_{C}(r)+U_{N}(r),
\]
where%
\[
U_{C}(r)=\left\{
\begin{array}
[c]{c}%
\dfrac{Z_{P}Z_{T}~e^{2}}{2R_{c}}\left(  3-\frac{r^{2}}{R_{c}^{2}}\right)
,\qquad\mathrm{for}\text{ }r<R_{c}\\
\qquad\dfrac{Z_{P}Z_{T}~e^{2}}{r},\qquad\qquad\quad~~~~\mathrm{for}\text{
}r\geq R_{c}\ .
\end{array}
\right.
\]
and%
\[
U_{N}(r)=\frac{-V_{0}}{1+\exp\left[  \left(  r-R_{r}\right)  /a_{r}\right]
}+\frac{-W_{0}}{1+\exp\left[  \left(  r-R_{i}\right)  /a_{i}\right]  }%
\]
Above, \ $R_{C}\ =r_{0C}\ \left(  A_{P}^{1/3}+A_{T}^{1/3}\right)
,~R_{r}\ =r_{0r}\ \left(  A_{P}^{1/3}+A_{T}^{1/3}\right)  $ and$~R_{i}%
\ =r_{0i}\ \left(  A_{P}^{1/3}+A_{T}^{1/3}\right)  .$ In the above equations,
we use typical values for the parameters:%
\begin{equation}%
\begin{array}
[c]{cc}%
\mathcal{F}_{0}~=3.0~\mathrm{MeV}; & r_{0C}=1.4\ \mathrm{fm}\\
V_{0}=60\ \mathrm{MeV}; & W_{0}=60\ \mathrm{MeV}\\
r_{0r}=1.25\ \mathrm{fm} & r_{0i}=1.00\ \mathrm{fm}\\
a_{r}=0.60\ \mathrm{fm}\, & a_{i}=0.60\ \mathrm{fm}.
\end{array}
\label{parameters}%
\end{equation}

\bigskip

The reduction of the above equations to get the polarization potential
proceeds as in the previous section and we get for the Schr\"{o}dinger
equation for $\psi_{0}(\mathbf{r})$ the following%
\begin{equation}
\left[  E-K-U^{opt}(r)-U^{pol}(\mathbf{r})\right]  ~\psi_{0}(\mathbf{r}%
)=0,\label{Upol5}%
\end{equation}
with%
\begin{equation}
U^{pol}(\mathbf{r},\mathbf{r}^{\prime})=\mathcal{F}(\mathbf{r})~G_{1}%
^{(+)}(\mathbf{r},\mathbf{r}^{\prime})~\mathcal{F}(\mathbf{r}^{\prime
}).\label{Upol6}%
\end{equation}

In order to perform numerical calculations it is convenient to carry out the
usual partial-wave expansions. Since we are using scalar form factors, the
partial-wave projected \ polarization potential is given by%
\begin{equation}
U_{l}^{pol}(r,r^{\prime})=\mathcal{F}(r)~g_{1,l}^{(+)}(r,r^{\prime
})~\mathcal{F}(r^{\prime}).\label{Upol5_l}%
\end{equation}
The partial-wave projected optical \bigskip Green function in channel-1 can be
written \cite{Sat83}
\begin{equation}
g_{1,l}^{(+)}(r,r^{\prime})=-\frac{2\mu}{\hbar^{2}k_{1}}~e^{-i\delta_{l}%
}~w_{l}(k_{1}r_{<})\;~\mathcal{H}_{l}^{(+)}(k_{1}r_{>}).\label{green-opt}%
\end{equation}
Above, $\mu$ is the reduced mass of the projectile-target system, $k_{1}%
=\sqrt{2\mu E_{1}}/\hbar,$ $r_{<}~(r_{>})$ is the smaller (larger) of the
radial separations $r$ and $r^{\prime}$, $w_{l}(k_{1}r)$ is the regular
solution of the optical Schr\"{o}dinger equation (partial-wave projected
Eq.(\ref{proj4}) setting $\mathcal{F}(\mathbf{r})=0$) with the asymptotic form%
\begin{equation}
w_{l}(k_{1}r\rightarrow\infty)=\frac{i}{2}\left[  H_{l}^{(-)}(k_{1}r)-\bar
{S}_{l}~H_{l}^{(-)}(k_{1}r)\right]  ,\label{wl}%
\end{equation}
and $\delta_{l}$\ is the nuclear phase-shift. Above, $H_{l}^{(-)}(H_{l}%
^{(+)})~$is the Coulomb wave function with ingoing (outgoing) boundary
condition and $\bar{S}_{l}=\exp\left(  2i~\delta_{l}\right)  $ is the nuclear
S-matrix at the $l^{th}$ partial-wave. In Eq.(\ref{green-opt}), $\mathcal{H}%
_{l}^{(+)}(k_{1}r)$ is the solution of the same optical Schr\"{o}dinger
equation, but with a different asymptotic behavior. At large separations, it
is the outgoing wave%
\[
\mathcal{H}_{1}^{(+)}(k_{1}r\rightarrow\infty)=\mathrm{e}^{i\delta_{l}}%
~H_{1}^{(+)}(k_{1}r_{\infty}).
\]
To determine $\mathcal{H}_{1}^{(+)}$ at finite projectile-target separations,
the radial equation must be numerically integrated inwards, starting from a
large $r-$value where the above asymptotic form is valid. This can be easily
achieved using a negative mesh step in any conventional code for numerical integration.

The dispersion relation is clearly satisfied by Eq. (\ref{Upol5_l}). Is it
satisfied by the $l-\mathrm{dependent}$ potential ${\bar{U}}_{l}$ (inserting
Eq.(\ref{Upol5_l}) in Eq.(\ref{Upol-barl}))? We remind that the expression
used to get ${\bar{U}}_{l}$ is Eq.(\ref{TELLP}). One would expect the
dispersion relation to be satisfied by ${\bar{U}}_{l}$ only if $u_{l}(kr)$ is
real, which is certainly not the case. However, a remnant of the dispersion
relation should still be seen in ${\bar{U}}_{l}$ as has been shown over and
over again in the study of the Threshold Anomaly \cite{Sat91}. Not
withstanding the above reservations we will give below an account of our
calculation of the equivalent \textit{l}-independent potential.

\begin{figure}[h]
\centering
\begin{minipage}[h] {1\linewidth}
\centering
\subfigure[]{
\includegraphics[width=0.70\linewidth]{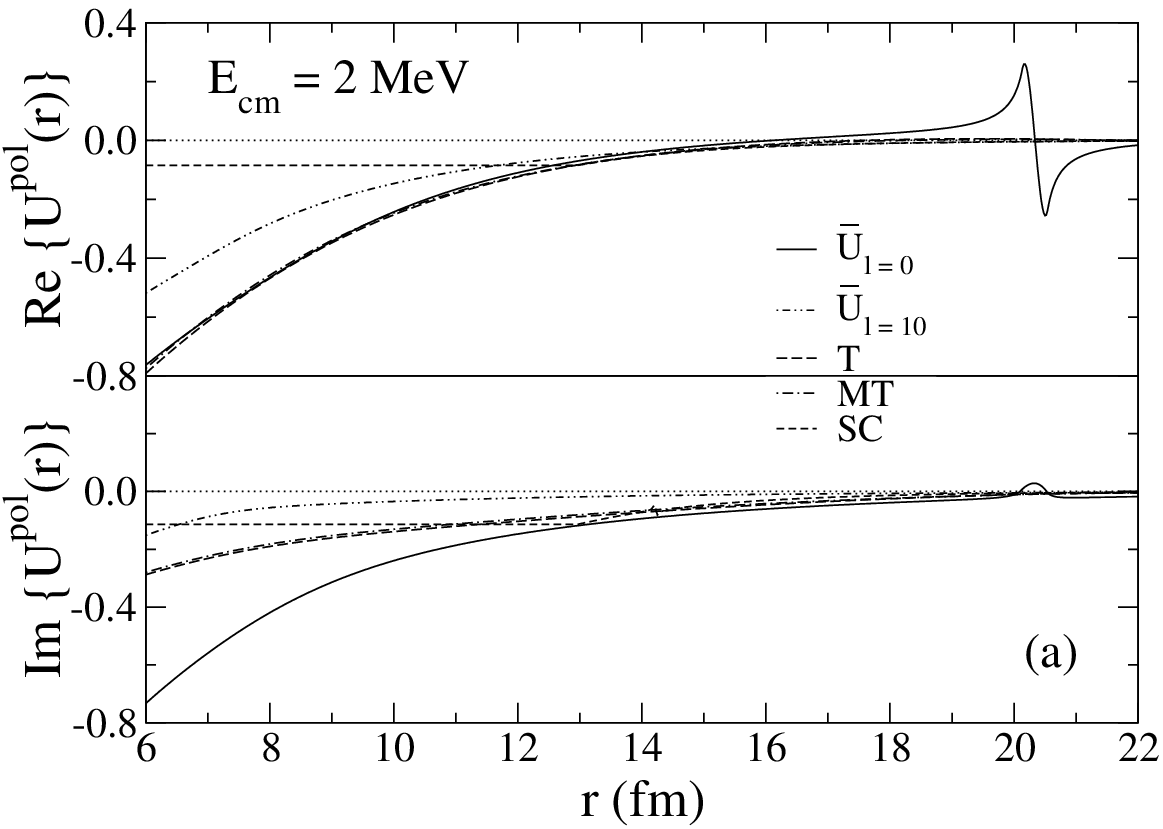}
}
\subfigure[]{
\includegraphics[width=0.70\linewidth]{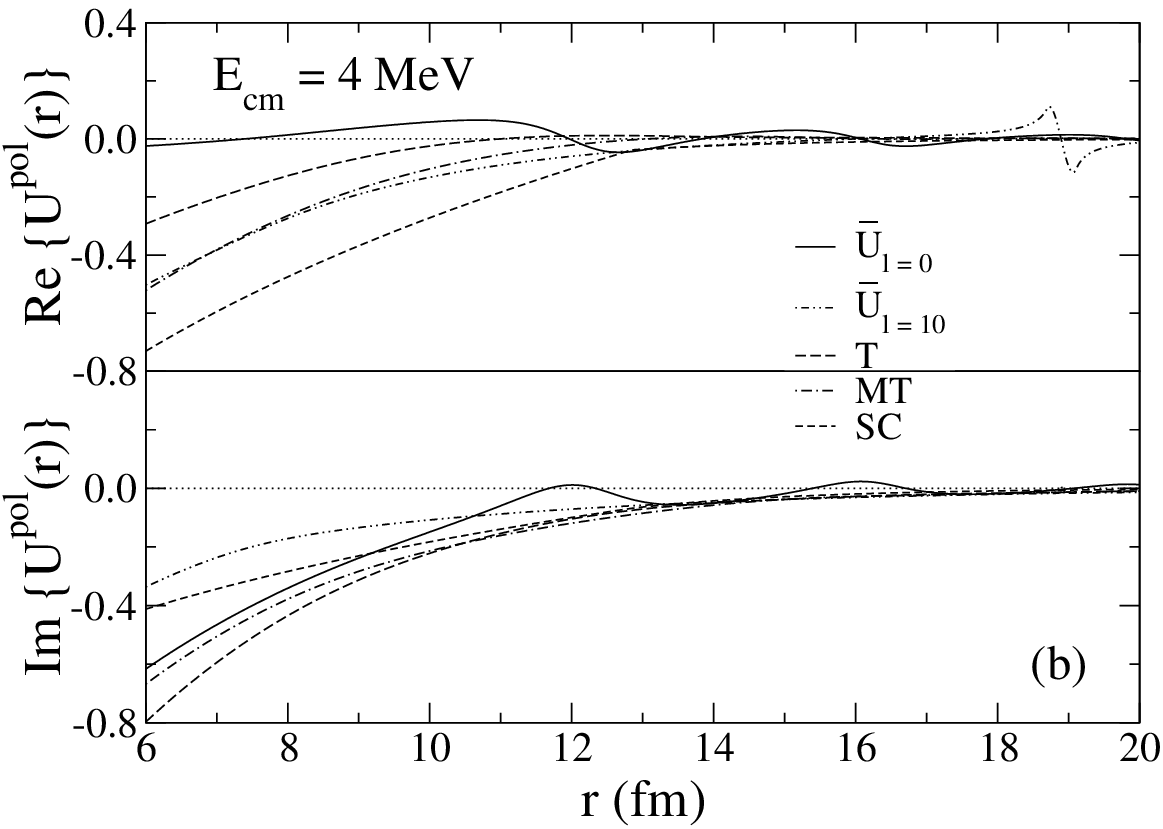}
}
\subfigure[]{
\includegraphics[width=0.70\linewidth]{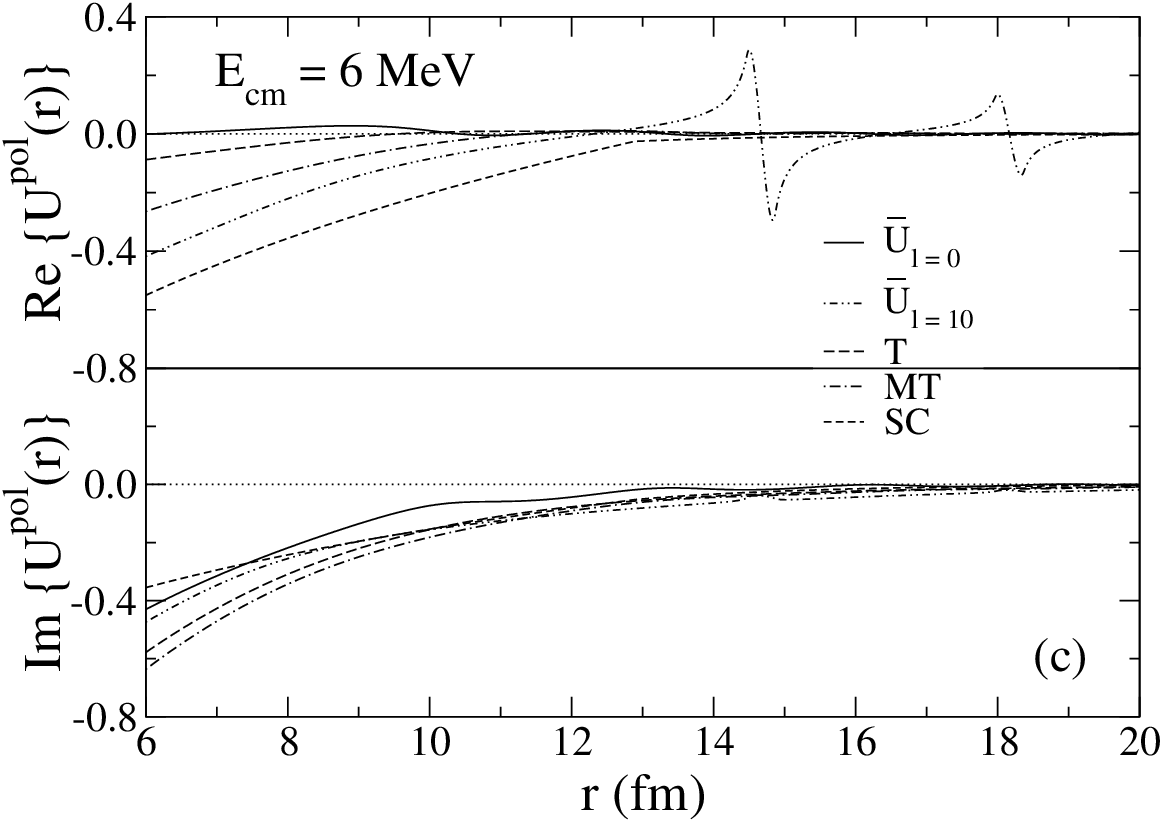}
}
\caption{{\small Real and imaginary parts of the polarization potentials
(in MeV) calculated using the different recipes discussed in the text.
T: Thompson, MT: modified Thompson, SC: Semiclassical, and the
$l-{\rm dependent}$ potentials for $l=0$ and $l=10$. The coupling
stength $\mathcal{F}_0$ is taken to be 3 MeV, a) $E = 2$ MeV, b) $E= 4$ MeV
and c) $E= 6$ MeV, for the system $^{11}$Li + $^{12}$C. See text for details.}}
\end{minipage}
\end{figure}

In figure 1 we show polarization potentials at the collision energies (a)
$E=2$ MeV, (b) $E=4$ MeV and (c) $E=6$ MeV. We present results obtained using
the Thompson (T) prescription (Eq.(\ref{VpolT})), the modified Thompson one
(MT) (Eq.(\ref{VpolTM})) and the semiclassical prescription (SC)
(Eq.(\ref{VpolSC} )). In the cases of the T and the MT potentials, we evaluate
the radial wave functions and the transmission coefficients solving the CC
equations and then carry out the $l-\mathrm{averages}$ of Eqs.(\ref{VpolT})
and (\ref{VpolTM}). Also shown are the approximate \textit{l}-dependent
potential ${\bar{U}}_{l}$ (Eq.(\ref{TELLP})) for $l=0$ and $l=10$ and the
optical potential employed in our two coupled channels. The singularities of
${\bar{U}}_{l}$, arising from the nodes of the radial wave functions, are
treated by the technique developed in Ref. \cite{VaV67}. We see clearly that
whereas the $\bar{U}_{l=0}$ is quite oscillatory, the T, MT and SC ones seem
to behave smoothly in so far as the real part of the potential is concerned.
Oscillations in $\bar{U}_{l}$ occur near the nodes of $w_{l}$. Since this wave
function is complex and their real and imaginary parts do not vanish
simultaneously, the polarization potential remains finite.

\begin{figure}[h]
\centering
\begin{minipage}[h] {1\linewidth}
\centering
\subfigure[]{
\includegraphics[width=0.70\linewidth]{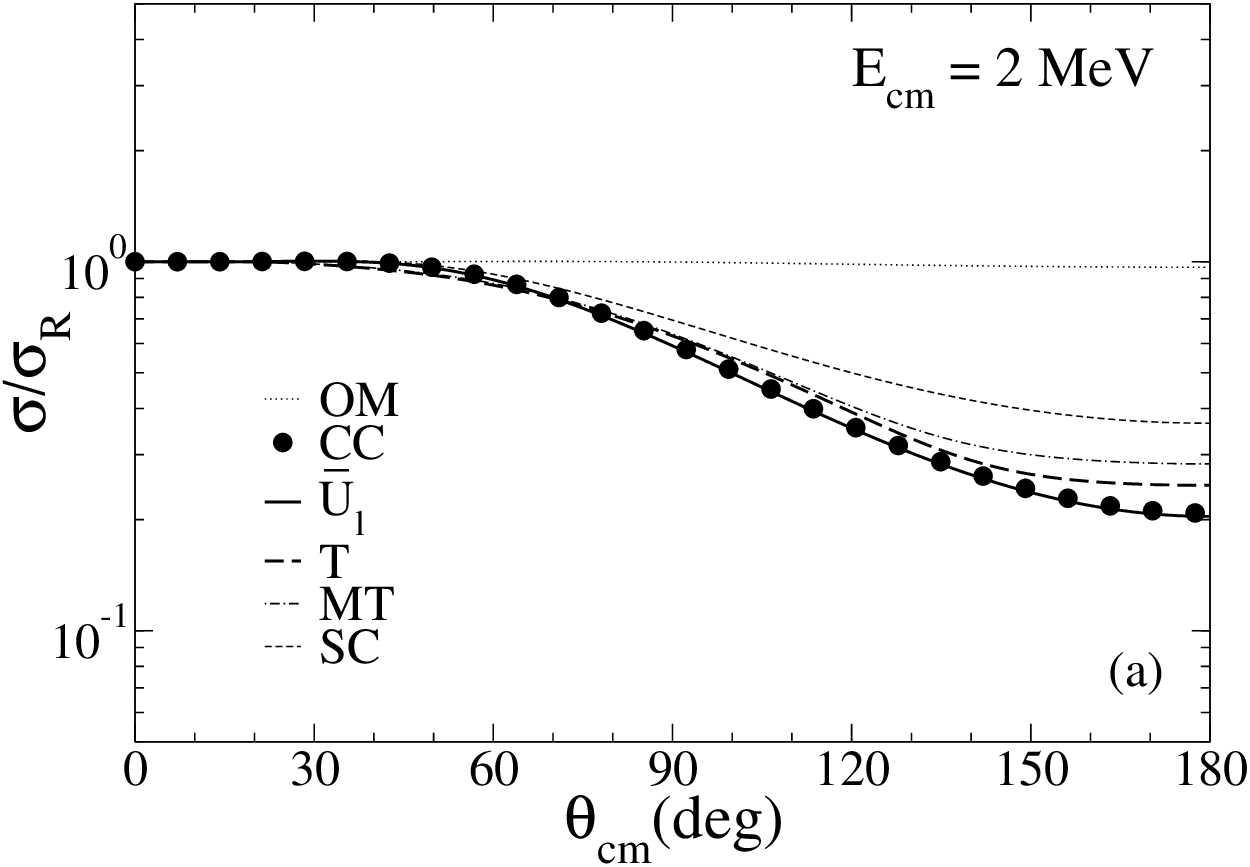}
}
\subfigure[]{
\includegraphics[width=0.70\linewidth]{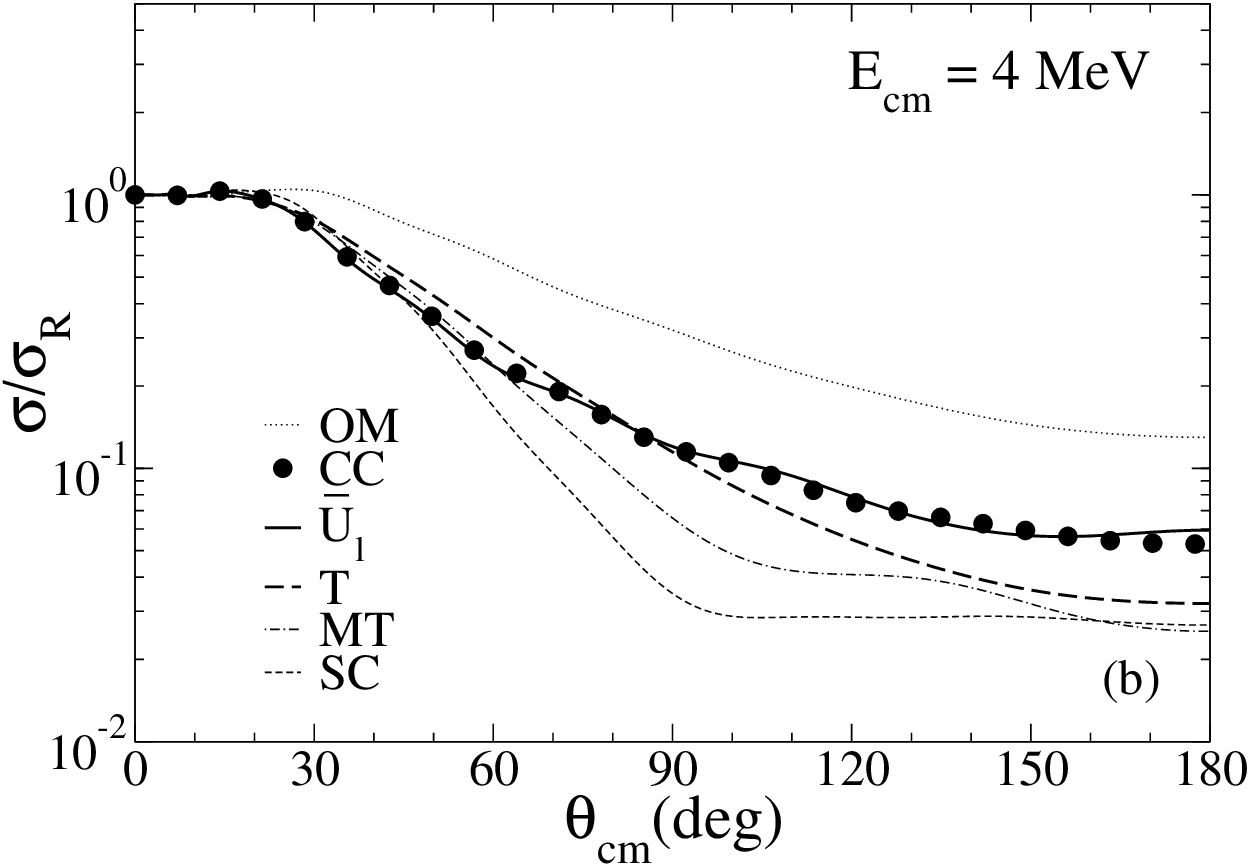}
}
\subfigure[]{
\includegraphics[width=0.70\linewidth]{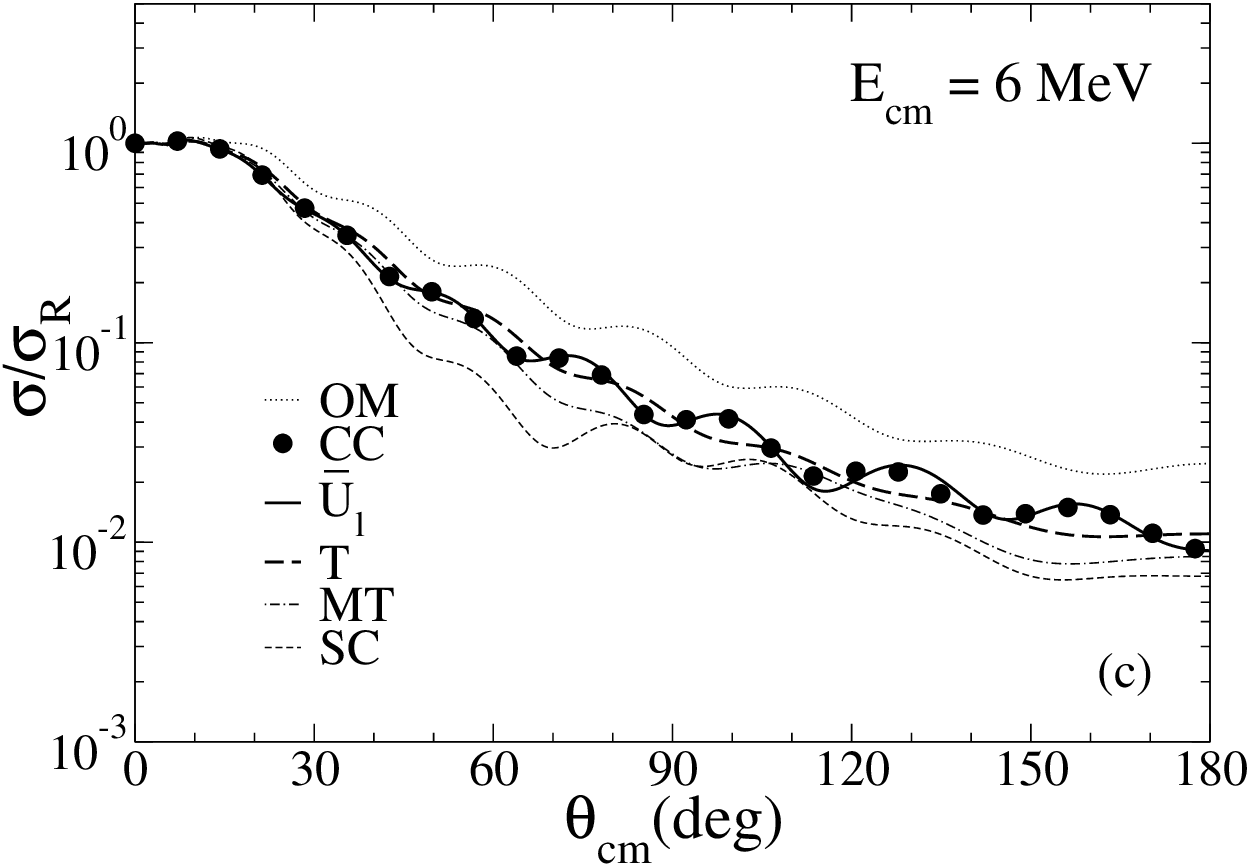}
}
\caption{{\small The ratio of the elastic angular distribution to Rutherford
calculated with CC (coupled channels), OM (optical model - no coupling) and
with the $l-{\rm dependent}$ polarization potential, $\bar{U}_l$. The other
symbols are the same as figure 1. See text for details.}}
\end{minipage}
\end{figure}

In figure 2 we show the results for the elastic scattering angular
distributions at $E=2,4,6$ MeV. The coupled channels results are shown as the
full circles. None of the \textit{l}-independent polarization potentials seems
to work very well. However, the T potential is better then the others. It is
close to the CC results, except for the collision energy of 6 MeV. In this case
the angular distribution obtained with this potential oscillates out of phase
with respect to the correspondint CC results. On the other hand, the approximate 
\textit{l}-dependent potential $\bar{U}_{l}$ reproduces accurately the CC results
at the three collision energies.

\begin{figure}[ptb]
\centering
\includegraphics[width=0.65\linewidth]{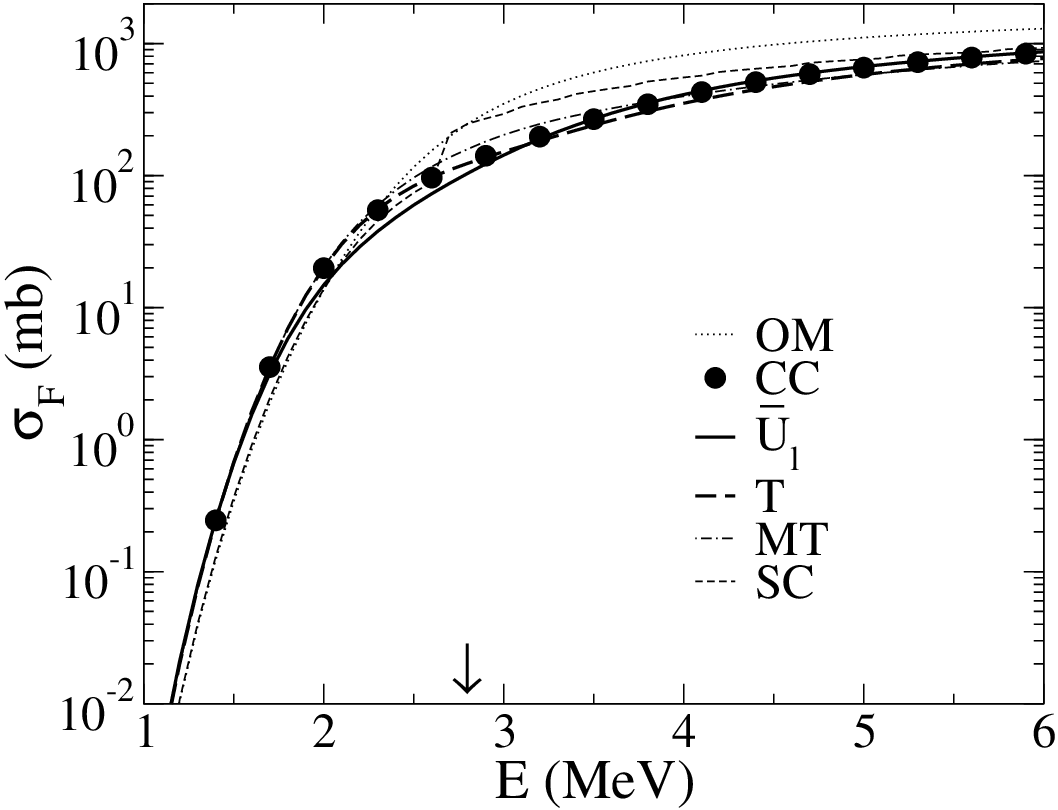}\caption{{\small The fusion
excitation function for the same cases as in figure 2. See text for details.}}%
\end{figure}

In figure 3 we show our results for the fusion cross section excitation
function, obtained by the relation%
\[
\sigma_{F}(E)=\frac{k}{E}\ \left\langle \psi_{0}\right\vert -\operatorname{Im}%
\{U^{opt}\}\left\vert \psi_{0}\right\rangle .
\]
The wave function $\psi_{0}$\ was calculated solving the Schr\"{o}dinger
equation for the elastic channel, including the optical and each of the above
discussed polarization potentials. Here, both ${\bar{U}}_{l}$ and the 
T-potential seem to work rather well over the energy range considered. The 
SC overshoots in the barrier region but otherwise it reproduces the CC 
calculation. The MT seems to be quite off, however on the average it works 
well too. 

\begin{figure}[ptb]
\centering
\includegraphics[width=0.65\linewidth]{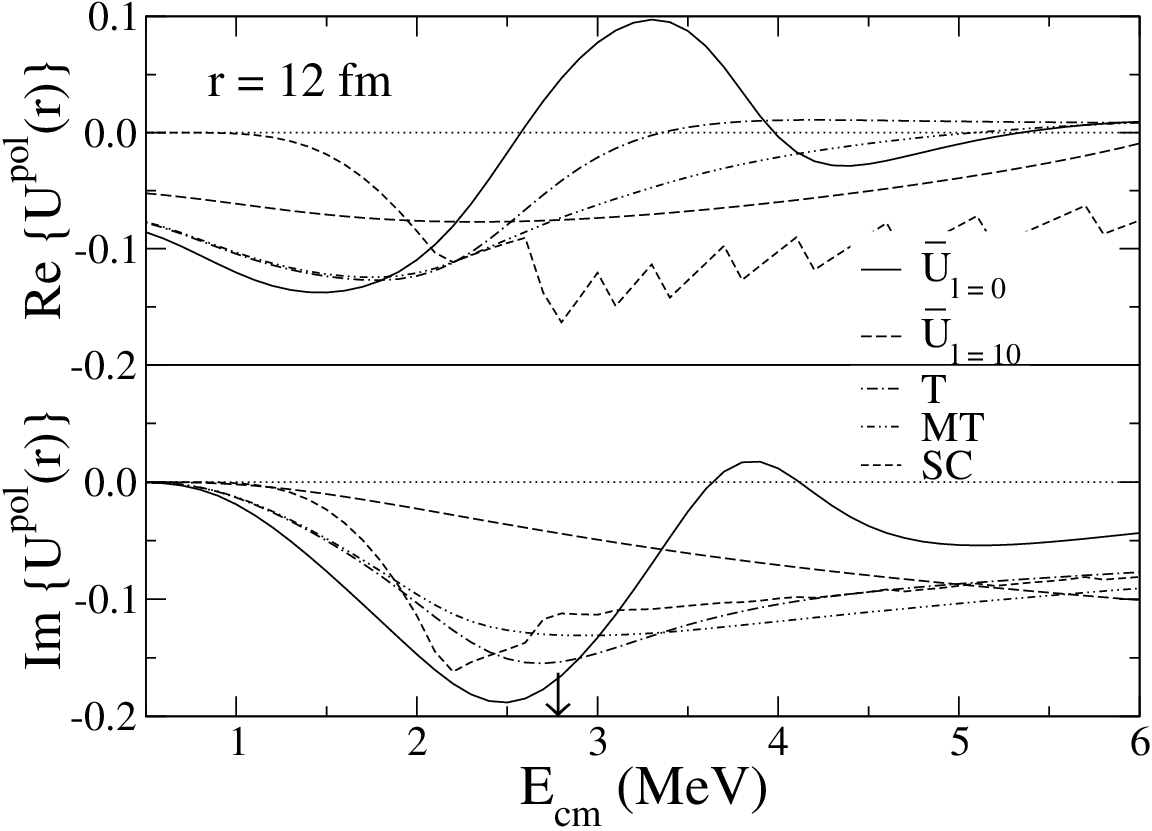}\caption{{\small The energy
dependence of the real and imaginay parts of the polarization potential (in
MeV), indicating the trend dicatated by the dispersion relation. See text for
details.}}%
\end{figure}

Finally, to be sure about the consistency of our calculation we have checked
the dispersion relation. We have fixed $r=12$ fm and plotted in figure 4 the
resulting behavior of the real and imaginary parts of the polarization
potential $vs$. $E$. It is clear that the ${\bar{U}}_{l}$ for $l=0$ shows the
general trend of what one would expect from the dispersion relation: maxima in
the real part accompanied by sharp variations in the imaginary part. Similar
behavior is found for the T potential. Note the unphysical sharp oscillations
of the SC potentials. They arise from the fact that this potential can only be
evaluated over a sparse mesh of r-values,correponding to the turning points at
each angular momentum. Thus the T and ${\bar{U}}_{l}$ potentials exhibit the
threshold anomaly \cite{Sat91}. In particular, the T potential shows a maximum
in the imaginary part at $E=3.9$ MeV, whereas the real part shows a maximum at
$E=3.5$ MeV. This behavior of the T-potential is in line with the breakup
threshold anomaly of Ref. \cite{HSL06}.

\section{Conclusions}

In this work we have investigated $l-\mathrm{independent}$ polarization
potentials in a schematic two-channel model, in which the range of the
polarization potentials is equivalent to that associated with the breakup
channel. We used several prescriptions to derive $l-\mathrm{independent}$
polarization potential and found that none could reproduce satisfactorily the
results of coupled channel calculations. This conclusion is consistent with
that of realistic CDCC calculations \cite{LuN07} using the Thompson
prescription. On the other hand, we have shown that an approximate
$l-\mathrm{dependent}$ potential, obtained using the unperturbed Green's
function in Feshbach theory, gave reasonable descriptions of elastic angular
distribution at energies above the barrier.

\section{Appendix}

We give below, a qualitative justification of the adopted parametrization of
$\mathcal{F}(r).$\ Since the diffusivity of the potentials is very small as
compared to $\alpha,R_{1}$ and $R_{2},$ we set $a_{r}\simeq0.$ In this way,
the Woods-Saxon potentials take the forms of the step functions,
\begin{equation}
U_{fT}(r_{fT})\simeq-V_{0}~\Theta\left(  \left\vert \mathbf{r}+\gamma
\mathbf{x}\right\vert -R_{T}\right)  . \label{UCTapprox}%
\end{equation}
Above, $f=f_{1}\ \mathrm{or} \ f_{2}$ and $\gamma=\gamma_{1}\ \mathrm{or}%
\ \gamma_{2}$.

At large projectile-target separations, $r>R_{P}+R_{T},$ only $x>R_{P}$
contributes to the integral of Eq.(\ref{Form2}). Since we are neglecting
angular momenta, in this range the ground state wave function depends only on
the radial and has an exponential form. That is,
\[
\varphi_{0}(x)\propto~\exp\left(  -\frac{x}{\alpha}\right)  ,
\]
with $\alpha$ given by Eq.(\ref{alpha}). Since we are adopting the sudden
approximation, the wave number of the state $\varphi_{1}$\ is supposed to be
vanishingly small. Therefore, this wave function is constant within the
integral. The contribution from the potential to the form factor takes the
form\
\begin{equation}
\mathcal{F}(r)\propto\int_{0}^{\infty}dx~\exp\left(  -\frac{x}{\alpha}\right)
~\int_{-1}^{1}\Theta\left(  \left\vert \mathbf{r}+\gamma\mathbf{x}\right\vert
-R_{T}\right)  ~dt,\label{Form5}%
\end{equation}
where $t$ stands for the cosine of the angle between the $\mathbf{x}$ and the
$\mathbf{r}$ vectors. The main contributions to this integral comes from
$\gamma\mathbf{x}$ anti-parallel to $\mathbf{r}$. We than replace%
\[
\Theta\left(  \left\vert \mathbf{r}+\gamma\mathbf{x}\right\vert -R_{T}\right)
\rightarrow\Theta\left(  r-\left\vert \gamma\right\vert x-R_{T}\right)
\]
and assume that the integration over $t$ does not depend strongly on $r,$
leading only to a renormalization of the strength of the form factor. The
integral of Eq.(\ref{Form5}) then becomes%
\[
\mathcal{F}(r)\propto\int_{\left(  r-R_{T}\right)  /\left\vert \gamma
\right\vert }^{\infty}dx~\exp\left(  -\frac{x}{\alpha}\right)  ~\propto
\exp\left[  -\frac{r}{\left\vert \gamma\right\vert ~\alpha}\right]  .
\]
Of course, the form factor will be dominated by the contribution from the
di-neutron, which has a longer range ($\left\vert \gamma_{1}\right\vert
>\left\vert \gamma_{2}\right\vert $). Denoting by $\mathcal{F}_{0}$ the
constant of proportionality, we can write%
\[
\mathcal{F}(r)\simeq\mathcal{F}_{0}~\exp\left(  -\frac{r}{\gamma_{1}\alpha
}\right)  ,
\]
which is adopted in the present calculation.

\bigskip

This work was supported in part by CNPq and the MCT/FINEP/CNPq(PRONEX) under
contract no. 41.96.0886.00. M.S.H. and W.H.Z.C. acknowledge support from the
FAPESP.

\bigskip\pagebreak

\end{document}